\documentclass[]{opennovelty}
\usepackage{fix-cm}
\usepackage{helvet}           

\usepackage{nicefrac}         
\usepackage{siunitx}         

\usepackage{longtable}
\usepackage{booktabs}

\usepackage{tabularx}         
\usepackage{array}           
\usepackage{makecell}         
\usepackage{threeparttable}

\usepackage{wrapfig}          
\usepackage{float}            
\usepackage[export]{adjustbox}

\usepackage{tikz}             
\usepackage{pgfplots}         
\usepackage{pgf-pie}

\usepackage{listings}        
\usepackage{ragged2e}        
\usepackage{comment}

\usepackage{pifont}           
\usepackage{fontawesome}

\usepackage{enumitem}        
\usepackage{titletoc}         
\usepackage{minitoc}          
\usepackage[toc,page,header]{appendix}

\usepackage{url}              
\usepackage[hang,flushmargin]{footmisc}  
\pgfplotsset{compat=1.18}

\definecolor{lightblue}{RGB}{200, 230, 255}  
\definecolor{headerblue}{RGB}{150, 200, 255}

\newcounter{examplebox}
\renewcommand{\theexamplebox}{\arabic{examplebox}}
\makeatletter
\newcommand\blfootnote[1]{%
  \begingroup
  \renewcommand\thefootnote{}\footnote{#1}%
  \addtocounter{footnote}{-1}%
  \endgroup
}
\makeatother

\title{%
  \begin{minipage}[c]{0.08\textwidth}
    \includegraphics[height=2.8em]{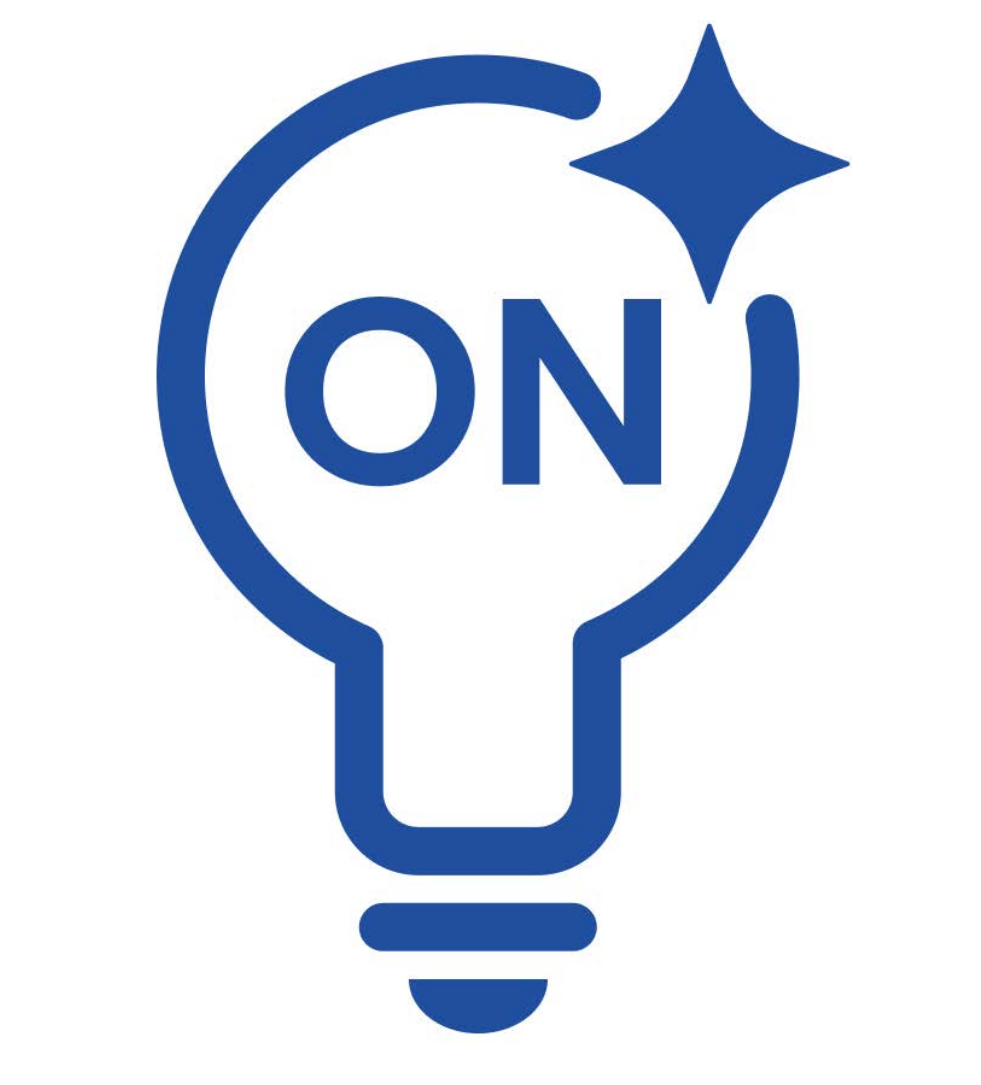} 
  \end{minipage}%
  \hspace{0.8em}%
  \begin{minipage}[c]{0.88\textwidth}
    \centering
    \textsc{OpenNovelty}: An LLM-powered Agentic System\\[0.3em]
    for Verifiable Scholarly Novelty Assessment
  \end{minipage}
}

\author{
    Ming Zhang\textsuperscript{1*$\dagger$}, 
    Kexin Tan\textsuperscript{1*},   
    Yueyuan Huang\textsuperscript{1*},
    Yujiong Shen\textsuperscript{1}, \\
    Chunchun Ma\textsuperscript{2},
    Li Ju\textsuperscript{2}, 
    Xinran Zhang\textsuperscript{3},
    Yuhui Wang\textsuperscript{1},
    Wenqing Jing\textsuperscript{1},\\
    Jingyi Deng\textsuperscript{1},
    Huayu Sha\textsuperscript{1},
    Binze Hu\textsuperscript{1}, 
    Jingqi Tong\textsuperscript{1},
    Changhao Jiang\textsuperscript{1},\\
    Yage Geng\textsuperscript{2},
    Yuankai Ying\textsuperscript{1,2},
    Yue Zhang\textsuperscript{2},
    Zhangyue Yin\textsuperscript{1},
    Zhiheng Xi\textsuperscript{1},\\
    Shihan Dou\textsuperscript{1},
    Tao Gui\textsuperscript{1},
    Qi Zhang\textsuperscript{1,2$\dagger$},
    Xuanjing Huang\textsuperscript{1}
}

\affiliation[1]{\mbox{Fudan University}}
\affiliation[2]{\mbox{WisPaper.AI}}
\affiliation[3]{\mbox{Claremont McKenna College}}

\abstract{Evaluating novelty is critical yet challenging in peer review, as reviewers must assess submissions against a vast, rapidly evolving literature. 
This report presents \textsc{OpenNovelty}, an LLM-powered agentic system for transparent, evidence-based novelty analysis. 
The system operates through four phases: (1) extracting the core task and contribution claims to generate retrieval queries; (2) retrieving relevant prior work based on extracted queries via semantic search engine; (3) constructing a hierarchical taxonomy of core-task-related work and performing contribution-level full-text comparisons against each contribution; and (4) synthesizing all analyses into a structured novelty report with explicit citations and evidence snippets. 
Unlike naive LLM-based approaches, \textsc{OpenNovelty} grounds all assessments in retrieved real papers, ensuring verifiable judgments. We deploy our system on 500+ ICLR 2026 submissions with all reports publicly available on our website, and preliminary analysis suggests it can identify relevant prior work, including closely related papers that authors may overlook.
\textsc{OpenNovelty} aims to empower the research community with a scalable tool that promotes fair, consistent, and evidence-backed peer review.

}

\correspondence{\email{mingzhang23@m.fudan.edu.cn},  \email{qz@fudan.edu.cn}}
\checkdata[Website]{\url{https://www.opennovelty.org}}

\begin{document}
\maketitle
\blfootnote{$^*$Equal Contribution.}
\blfootnote{$^\dagger$Corresponding authors.}


\vspace{-1.5em}

\begin{figure}[t]
\centering
\includegraphics[width=\textwidth]{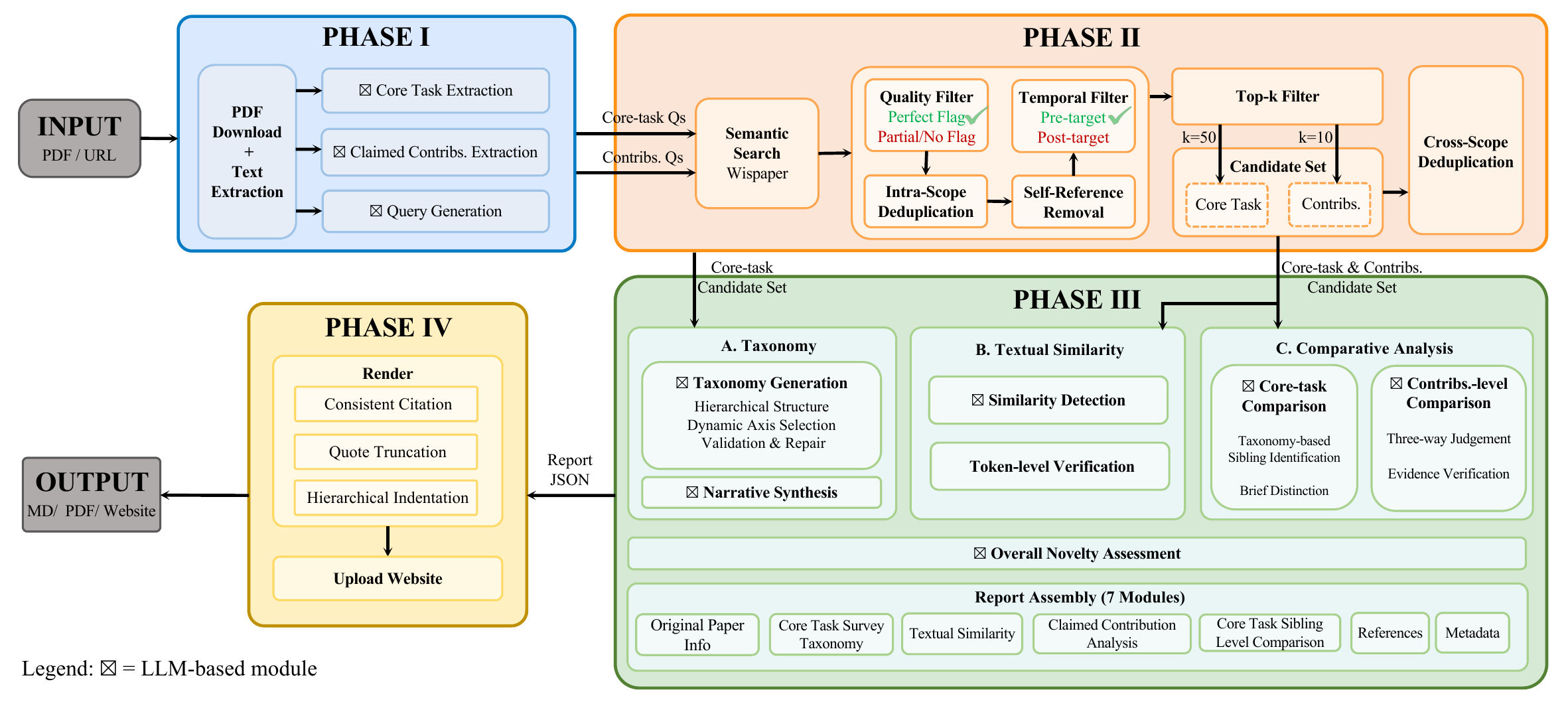}
\caption{Overview of the \textsc{OpenNovelty} framework. \textbf{Phase~I} extracts the core task and claimed contributions and generates expanded queries. \textbf{Phase~II} retrieves and filters candidate prior work. \textbf{Phase~III} constructs a taxonomy and performs evidence-verified comparisons. \textbf{Phase~IV} renders the final novelty report from structured outputs.}
\label{fig:framework}
\end{figure}

\section{Introduction}
\label{sec:intro}

In recent years, academic publications have grown exponentially~\cite{santo2018science}. In Artificial Intelligence alone, the \texttt{cs.AI} and \texttt{cs.LG} categories on arXiv receive tens of thousands of new papers annually, while submissions to top-tier conferences (NeurIPS, ICLR, ICML) continue to hit record highs~\cite{lixin4ever_conference_acceptance_rate}. This ``publication explosion'' places unprecedented pressure on the peer review system~\cite{huang2025dual, latona2024aireviewlotterywidespread}.

The burden on reviewers has intensified significantly. 
A single reviewer is often required to evaluate multiple papers within a limited timeframe, with each review demanding a comprehensive understanding of frontier work in the relevant field. 
However, in reality, many reviewers lack the time and energy to conduct thorough, fair assessments for every submission.
In extreme cases, some reviewers provide feedback without carefully reading the full text. 
Furthermore, the academic community has increasingly voiced concerns about reviewers using AI-generated feedback without proper verification, a practice that threatens the integrity of the peer review process~\cite{iclr2026_llm_response, liang2024monitoring}.

Among various evaluation dimensions, \textbf{novelty} is widely regarded as a critical determinant of acceptance. 
However, accurately assessing novelty remains a formidable challenge due to the vast scale of literature, the difficulty of verifying claims through fine-grained analysis, and the subjectivity inherent in reviewers' judgments.
While Large Language Models (LLMs) have emerged as a promising direction for assisting academic review~\cite{zhuang2025large, lin2023automated, lin2023a, zhang2025evolving, xu2025can, wu2025automated, lin2025evaluating, shahid2025literature}, existing work faces significant limitations: naive LLM-based approaches may hallucinate non-existent references when relying solely on parametric knowledge~\cite{lin2023a, zhang2025evolving, xu2025can};
existing RAG-based methods ~\cite{lin2025evaluating, wu2025automated} compare only titles and abstracts, missing critical technical details; and most approaches are constrained by context windows, lacking systematic organization of related work~\cite{shahid2025literature}.

To address these challenges, we introduce \textbf{\textsc{OpenNovelty}}, an LLM-powered agentic system designed to provide transparent, traceable, and evidence-based novelty analysis for large-scale submissions. Unlike existing methods, the core design philosophy of \textsc{OpenNovelty} is to make \emph{\textbf{Novelty Verifiable}}:

\begin{quote}
\textit{``We do not rely on parametric knowledge in LLMs; instead, we retrieve real papers and perform contribution-level full-text comparisons to ensure \textbf{every judgment} is grounded in evidence.''}
\end{quote}

\textsc{OpenNovelty} operates through a four-phase framework:

\begin{itemize}
    \item \textbf{Phase~I: Information Extraction} — Extracts the core task and contribution claims from the target paper, and generates semantic queries for subsequent retrieval.

    \item \textbf{Phase~II: Paper Retrieval} — Based on the extracted queries, retrieves relevant prior work via a semantic search engine (\textsc{Wispaper}~\cite{ju2025wispaperaischolarsearch}) and employs multi-layer filtering to select high-quality candidates.
      
    \item \textbf{Phase~III: Analysis \& Synthesis} — Based on the extracted claims and retrieved papers, constructs a hierarchical taxonomy of related work while performing full-text comparisons to verify each contribution claim.
    
    \item \textbf{Phase~IV: Report Generation} — Synthesizes all preceding analyses into a structured novelty report with explicit citations and evidence snippets, ensuring all judgments are verifiable and traceable.
\end{itemize}

\noindent Technical details for each phase are provided in Section~\ref{sec:opennovelty}.

Additionally, we deployed \textsc{OpenNovelty} to analyze 500+ highly-rated submissions to ICLR 2026, with all novelty reports publicly available on our website\footnote{\url{https://www.opennovelty.org}}.
Preliminary analysis suggests that the system can identify relevant prior work, including closely related papers that authors may overlook. We plan to scale this analysis to over 2,000 submissions in subsequent phases.

Our main contributions are summarized as follows:

\begin{itemize}
    \item We propose a novelty analysis framework that integrates contribution extraction, semantic retrieval, LLM-driven taxonomy construction, and contribution-level comparison into a fully automated pipeline. The hierarchical taxonomy provides reviewers with structured context to understand each paper's positioning within the research field.
    
    \item We ground all novelty judgments in retrieved real papers, with each judgment accompanied by explicit citations and evidence snippets, effectively avoiding hallucination issues common in naive LLM-based approaches.
    
    \item We deploy \textsc{OpenNovelty} on 500+ ICLR 2026 submissions and publicly release all reports on our website, providing accessible and transparent novelty analysis for the research community.
\end{itemize}
\section{\textsc{OpenNovelty}}
\label{sec:opennovelty}
In this section, we detail the four core phases of \textsc{OpenNovelty}. An overview of our framework is illustrated in Figure~\ref{fig:framework}. To illustrate the workflow of each phase, we use a recent paper from arXiv on reinforcement learning for LLM agents~\cite{xi2025agentgym} as a running example throughout this section.

\subsection{Phase~I: Information Extraction}
\label{sec:phase1}

The first phase extracts key information from the target paper and generates semantic queries for subsequent retrieval. Specifically, this phase involves two steps: (1) extracting the \textbf{core task} and \textbf{contribution claims}; and (2) generating \textbf{semantic queries with variants} for retrieving relevant prior work. All extraction tasks are performed using \texttt{claude-sonnet-4-5-20250929}~\cite{anthropic_claude_sonnet_4_5_20250930} with carefully engineered prompts in a zero-shot paradigm. 

\subsubsection{Core Task and Contribution Extraction}

\paragraph{Core Task.}
We extract the main problem or challenge that the paper addresses, expressed as a single phrase of 5--15 words using abstract field terminology (e.g., ``accelerating diffusion model inference'') rather than specific model names introduced in the paper. This abstraction ensures that the generated queries can match a broader range of related work.
The prompt template used for core task extraction is provided in Appendix~\ref{app:phase1_prompts}, Table~\ref{tab:app_prompt_core_task}.

\paragraph{Claimed Contributions.}
We extract the key contributions claimed by the authors, including novel methods, architectures, algorithms, datasets, benchmarks, and theoretical formalizations. Pure experimental results and performance numbers are explicitly excluded. Each contribution is represented as a structured object containing four fields: (1) a concise \texttt{name} of at most 15 words; (2) verbatim \texttt{author\_claim\_text} of at most 40 words for accurate attribution; (3) a normalized \texttt{description} of at most 60 words for query generation; and (4) a \texttt{source\_hint} for traceability. The LLM locates contribution statements using cue phrases such as ``We propose'' and ``Our contributions are'', focusing on the title, abstract, introduction, and conclusion sections.
The prompt template for contribution claim extraction is detailed in Appendix~\ref{app:phase1_prompts}, Table~\ref{tab:app_prompt_contribution_extraction}.

\subsubsection{Query Generation with Semantic Expansion}

Based on the extracted content, we generate semantic queries for Phase~II retrieval. We adopt a query expansion mechanism that produces multiple semantically equivalent variants~\cite{manning2008introduction}. The prompt templates for primary query generation and semantic variant expansion are provided in Appendix~\ref{app:phase1_prompts}, Tables~\ref{tab:app_prompt_primary_query} and~\ref{tab:app_prompt_query_variants}.

\paragraph{Generation Process.}
For each extracted item (core task or contribution), we first generate a primary query synthesized from the extracted fields while preserving key terminology. We then generate two semantic variants, which are paraphrases that use alternative academic terminology and standard abbreviations (e.g., ``RL'' for ``reinforcement learning''). Contribution queries follow the format ``Find papers about [topic]'' with a soft constraint of 5--15 words and a hard limit of 25 words, while core task queries are expressed directly as short phrases without the search prefix. Example~\ref{box:query_example} illustrates a typical query generation output.

\refstepcounter{examplebox}
\begin{tcolorbox}[
    colback=headerblue!10, 
    colframe=headerblue!50, 
    title={Example~\theexamplebox: Query Generation from Extraction Results},
    fonttitle=\bfseries,
    breakable,
    enhanced,
    float,                 
    floatplacement={!t},
]
\label{box:query_example}
\small

\textbf{A. Core Task Query} (short phrase without search prefix)

\vspace{0.3em}
\textbf{Extracted Core Task:}
\begin{itemize}[leftmargin=*, nosep]
    \item \texttt{text}: ``training LLM agents for long-horizon decision making via multi-turn reinforcement learning''
\end{itemize}

\vspace{0.3em}
\textbf{Generated Queries:}
\begin{enumerate}[leftmargin=*, nosep]
    \item \textbf{Primary:} ``training LLM agents for long-horizon decision making via multi-turn reinforcement learning''
    \item \textbf{Variant 1:} ``multi-step RL for training large language model agents in long-term decision tasks''
    \item \textbf{Variant 2:} ``reinforcement learning of LLM agents across extended multi-turn decision horizons''
\end{enumerate}

\vspace{0.8em}

\textbf{B. Contribution Query} (with ``Find papers about'' prefix)

\vspace{0.3em}
\textbf{Extracted Contribution:}
\begin{itemize}[leftmargin=*, nosep]
    \item \texttt{name}: ``AgentGym-RL framework for multi-turn RL-based agent training''
    \item \texttt{description}: ``A unified reinforcement learning framework with modular architecture that supports mainstream RL algorithms across diverse scenarios including web navigation and embodied tasks.''
\end{itemize}

\vspace{0.3em}
\textbf{Generated Queries:}
\begin{enumerate}[leftmargin=*, nosep]
    \item \textbf{Primary:} ``Find papers about reinforcement learning frameworks for training agents in multi-turn decision-making tasks''
    \item \textbf{Variant 1:} ``Find papers about RL systems for learning policies in multi-step sequential decision problems''
    \item \textbf{Variant 2:} ``Find papers about reinforcement learning methods for agent training in long-horizon interactive tasks''
\end{enumerate}

\end{tcolorbox}

\paragraph{Output Statistics.}
Each paper produces 6--12 queries in total: 3 queries for the core task (1 primary plus 2 variants) and 3--9 queries for contributions (1--3 contributions multiplied by 3 queries each).

\vspace{1em}
\subsubsection{Implementation and Output}
Phase~I involves several technical considerations: \textbf{zero-shot prompt engineering} for extracting the core task and author-stated contributions, \textbf{structured output validation} with parsing fallbacks and constraint enforcement, \textbf{query synthesis with semantic variants} under explicit format and length rules, and \textbf{publication date inference} to support temporal filtering in Phase~II. For long documents, we truncate the paper text at the ``References'' section with a hard limit of 200K characters. Appendix~\ref{app:phase1} provides the corresponding specifications, including output field definitions, temperature settings, prompt design principles, validation and fallback mechanisms, and date inference rules.

The rationale for adopting a \textbf{zero-shot paradigm} and \textbf{query expansion strategy} is discussed in Section~\ref{sec:discuss_extraction}; limitations regarding \textbf{mathematical formulas} and \textbf{visual content} extraction are addressed in Section~\ref{sec:limit_content}.

The outputs of Phase~I include the core task, contribution claims, and 6--12 expanded queries. These outputs serve as inputs for \textbf{Phase~II} and \textbf{Phase~III}.

\subsection{Phase~II: Paper Retrieval}
\label{sec:phase2}

Phase~II retrieves relevant prior work based on the queries generated in Phase~I. We adopt a \textit{broad recall, multi-layer filtering} strategy: the semantic search engine retrieves all potentially relevant papers (typically hundreds to thousands per submission), which are then distilled through sequential filtering layers to produce high-quality candidates for subsequent analysis. 

\subsubsection{Semantic Search}

We use \textsc{Wispaper}~\cite{ju2025wispaperaischolarsearch} as our semantic search engine, which is optimized for academic paper retrieval. Phase~II directly uses the natural language queries generated by Phase~I without any post-processing. Queries are sent to the search engine exactly as generated, without keyword concatenation or Boolean logic transformation. This design preserves the semantic integrity of LLM-generated queries and leverages \textsc{Wispaper}'s natural language understanding capabilities.

\paragraph{Execution Strategy.}
All 6--12 queries per paper are executed using a thread pool with configurable concurrency (default: 1 to respect API rate limits; configurable for high-throughput scenarios).

\paragraph{Quality Flag Assignment.}
For each retrieved paper, we compute a quality flag (\texttt{perfect}, \texttt{partial}, or \texttt{no}) based on \textsc{Wispaper}'s verification verdict. Only papers marked as \texttt{perfect} proceed to subsequent filtering layers.

\subsubsection{Multi-layer Filtering}
Raw retrieval results may contain hundreds to thousands of papers. We apply scope-specific filtering pipelines---separately for core task and contribution queries---followed by cross-scope deduplication to produce a high-quality candidate set. Importantly, we rely on semantic relevance signals rather than citation counts or venue prestige.

\paragraph{Core Task Filtering.}

For the 3 core task queries (1 primary plus 2 variants), we apply sequential filtering layers: (1) \textbf{quality flag filtering} retains only papers marked as \texttt{perfect}, indicating high semantic relevance, typically reducing counts by approximately 70--80\%; (2) \textbf{intra-scope deduplication} removes papers retrieved by multiple queries within this scope using canonical identifier matching (MD5 hash of normalized title), typically achieving a 20--50\% reduction depending on query overlap; (3) \textbf{Top-K selection} ranks remaining candidates by relevance score and selects up to \textbf{50 papers} to ensure broad coverage of related work.

\paragraph{Contribution Filtering.}

For each of the 1--3 contributions, we apply the same filtering pipeline to its 3 queries: quality flag filtering followed by Top-K selection of up to \textbf{10 papers} per contribution. Since contribution queries are more focused, intra-scope deduplication within each contribution typically yields minimal reduction. Together, contributions produce 10--30 candidate papers.

\paragraph{Cross-scope Deduplication.}

After Top-K selection for both scopes, we merge the core task candidates (up to 50 papers) and contribution candidates (10--30 papers) into a unified candidate set. We then perform cross-scope deduplication by identifying papers with identical canonical identifiers that appear in both scopes, prioritizing the instance from the core task list with higher-quality metadata (DOI $>$ arXiv $>$ OpenReview $>$ title hash). This step typically removes 5--15\% of combined candidates. The final output contains 60--80 unique papers per submission.

\paragraph{Additional Filtering.}
Two additional filters are always applied during per-scope filtering, after quality filtering and before Top-K selection: 
(1) \textbf{self-reference removal} filters out the target paper itself using canonical identifier comparison, PDF URL matching, or direct title matching; 
(2) \textbf{temporal filtering} excludes papers published after the target paper to ensure fair novelty comparison.
In many cases these filters remove zero papers, but we keep them as mandatory steps for correctness.

Example~\ref{box:filtering_example} illustrates a typical filtering progression.

\refstepcounter{examplebox}
\begin{tcolorbox}[
    colback=headerblue!10, 
    colframe=headerblue!50, 
    title={Example~\theexamplebox: Filtering Progression},
    fonttitle=\bfseries,
    breakable,
    enhanced,
    float,                 
    floatplacement={!t},
]
\label{box:filtering_example}
\small
\textbf{Target Paper:} ``AgentGym-RL: Training LLM Agents for Long-Horizon Decision Making through Multi-Turn Reinforcement Learning''

\vspace{0.5em}
\textbf{A. Core Task Filtering} (3 queries)
\begin{itemize}[leftmargin=*, nosep]
    \item Raw retrieval: 774 papers
    \item After quality flag filtering: 210 papers ($-$72.9\%)
    \item After intra-scope deduplication: 163 papers ($-$22.4\%)
    \item Top-K selection (K=50): \textbf{50 papers}
\end{itemize}

\vspace{0.5em}
\textbf{B. Contribution Filtering} (9 queries across 3 contributions)
\begin{itemize}[leftmargin=*, nosep]
    \item Raw retrieval: 1,554 papers
    \item After quality flag filtering: 336 papers ($-$78.4\%)
    \item Top-K selection (K=10 per contribution): \textbf{30 papers}
\end{itemize}

\vspace{0.5em}
\textbf{C. Cross-scope Deduplication}
\begin{itemize}[leftmargin=*, nosep]
    \item Combined Top-K candidates: 50 + 30 = 80 papers
    \item After cross-scope deduplication: \textbf{73 unique papers} ($-$8.8\%)
\end{itemize}

\vspace{0.3em}
\textit{Overall reduction: 2,328 $\rightarrow$ 73 papers (96.9\% filtered)}\\
\textit{Note: In this example, self-reference and temporal filters remove zero papers, but the steps are still executed.}

\end{tcolorbox}

\subsubsection{Implementation and Output}
Phase~II involves several technical considerations: \textbf{fault-tolerant query execution} with automatic retry, token refresh, and graceful degradation under partial failures; \textbf{quality-flag computation} (\texttt{perfect}, \texttt{partial}, \texttt{no}) derived from \textsc{Wispaper}'s verification verdict to gate downstream filtering; and \textbf{multi-layer candidate filtering} including intra-scope and cross-scope deduplication, self-reference removal, temporal filtering, and Top-K selection. Appendix~\ref{app:phase2} provides the corresponding specifications, including filtering statistics, fault-tolerance mechanisms, and the quality-flag mapping rules. 

The rationale for adopting a \textbf{broad recall strategy}, prioritizing \textbf{semantic relevance over citation metrics}, and using \textbf{natural language queries} is discussed in Section~\ref{sec:discuss_retrieval}; limitations regarding \textbf{indexing scope} and \textbf{result interpretation} are addressed in Section~\ref{sec:limit_retrieval}.

Phase~II produces up to 50 core task candidates and up to 10 candidates per contribution, typically totaling 60--80 unique papers per submission. Core task candidates feed into taxonomy construction in \textbf{Phase~III}, while contribution candidates are used for contribution-level novelty verification.

\subsection{Phase~III: Analysis \& Synthesis}
\label{sec:phase3}
The third phase performs the core analytical tasks of \textsc{OpenNovelty}. Based on the candidates retrieved in Phase~II, this phase involves constructing a hierarchical \textbf{taxonomy} that organizes related work, performing unified \textbf{textual similarity detection}, and conducting \textbf{comparative analyses} across both core-task and contribution scopes. All analytical tasks are performed using \texttt{claude-sonnet-4-5-20250929}.

\subsubsection{Taxonomy Construction}

We construct a hierarchical taxonomy to organize the Top-50 core task papers retrieved in Phase~II. Unlike traditional clustering algorithms that produce distance-based splits without semantic labels, our LLM-driven approach generates interpretable category names and explicit scope definitions at every level. 

\paragraph{Hierarchical Structure.}
The taxonomy adopts a flexible hierarchical structure with a typical depth of 3--5 levels. The \textbf{root} node represents the overall research field derived from the core task. \textbf{Internal} nodes represent major methodological or thematic categories. \textbf{Leaf} nodes contain clusters of 2--7 semantically similar papers. We enforce MECE (Mutually Exclusive, Collectively Exhaustive) principles by requiring each node to include a \texttt{scope\_note} (a concise inclusion rule) and an \texttt{exclude\_note} (an exclusion rule specifying where excluded items belong), ensuring that every retrieved paper belongs to exactly one category.

\paragraph{Generation Process.}
The LLM receives the metadata (title and abstract) of all Top-50 papers along with the extracted core task, and generates the complete taxonomy in a single inference call. The prompt instructs the LLM to dynamically select the classification axis that best separates the papers. By default, it considers methodology first, then problem formulation, and finally study context---but only when these axes yield clear category boundaries. Each branch node must include both a \texttt{scope\_note} field (a concise inclusion rule) and an \texttt{exclude\_note} field (an exclusion rule specifying where excluded items belong) to enforce MECE constraints.
The prompt template for taxonomy construction is provided in Appendix~\ref{app:phase3_prompts}, Table~\ref{tab:app_prompt_taxonomy_construction}.

\paragraph{Validation and Repair.}
Generated taxonomies undergo automatic structural validation. We verify that every retrieved paper is assigned to exactly one leaf node (coverage check), that no paper appears in multiple categories (uniqueness check), and that all referenced paper IDs exist in the candidate set (hallucination check). For violations, we apply a two-stage repair strategy: (1) deterministic pre-processing removes extra IDs and deduplicates assignments, then (2) if coverage errors persist, we invoke a single LLM repair round that places missing papers into best-fit existing leaves based on their titles and abstracts. We adopt a strict policy: rather than creating artificial ``Unassigned'' categories to force invalid taxonomies to pass validation, we mark taxonomies that remain invalid after repair as \texttt{needs\_review} and allow them to proceed through the pipeline with diagnostic annotations 
for manual inspection. The prompt template for taxonomy repair is provided in Appendix~\ref{app:phase3_prompts}, Table~\ref{tab:app_prompt_taxonomy_repair}.

\subsubsection{Textual Similarity Detection}
Before performing comparative analysis, we detect potential textual overlap between papers, which may indicate duplicate submissions, undisclosed prior versions, or extensive unattributed reuse. The prompt template for textual similarity detection is provided in Appendix~\ref{app:phase3_prompts}, Table~\ref{tab:app_prompt_similarity_detection}.

\paragraph{Segment Identification.}
Each candidate paper (identified by its canonical ID) is analyzed exactly once to avoid redundant LLM calls. The LLM is instructed to identify \textbf{similarity segments}: contiguous passages of 30 or more words that exhibit high textual overlap between the target and candidate papers. For each identified segment, the system extracts the approximate location (section name) and content from both papers. Segments are classified into two categories: \texttt{Direct} indicates verbatim copying without quotation marks or citation; \texttt{Paraphrase} indicates rewording without changing the core meaning or providing attribution. Results are cached and merged into the comparison outputs at the end of Phase~III, before report generation.

\paragraph{Segment Verification.}
While the LLM provides initial segment identification, we verify each reported segment against the source texts using the same token-level anchor alignment algorithm as evidence verification (Appendix~\ref{app:evidence_verification}). Both the reported original and candidate text segments are normalized (lowercased, whitespace-collapsed) and verified against the full normalized texts of both papers. A segment is accepted only if \emph{both} quotes are successfully verified by the anchor-alignment verifier (with a confidence threshold of 0.6) in their respective source documents and the segment meets the 30-word minimum threshold. Each verified segment includes a brief rationale explaining the nature of the textual overlap. The final report presents these findings for human interpretation without automated judgment, as high similarity may have legitimate explanations such as extended versions or shared authors.

\subsubsection{Comparative Analysis}

We perform two types of comparative analysis based on candidates retrieved in Phase~II. The prompt templates for contribution-level comparison, core-task sibling distinction, and subtopic-level comparison are provided in Appendix~\ref{app:phase3_prompts}, Tables~\ref{tab:app_prompt_claim_comparison}, \ref{tab:app_prompt_core_task_distinction}, and~\ref{tab:app_prompt_subtopic_comparison}. Example~\ref{box:comparison_example} illustrates typical outputs for both types.

\paragraph{Core-Task Comparison.}
Core-task comparisons differentiate the target paper from its structural neighbors in the taxonomy. The comparison adapts to structural position: when sibling papers exist in the same taxonomy leaf, we generate individual paper-level distinctions; when no siblings exist but sibling subtopics are present, we perform categorical comparison; isolated papers with no immediate neighbors are logged without comparison (see Appendix~\ref{app:comparison_paths}). For each comparison, the output includes a \texttt{brief\_comparison} field summarizing methodological or conceptual distinctions, and an \texttt{is\_duplicate\_variant} flag indicating whether the two papers appear to be versions of the same work.

\paragraph{Contribution-level Comparison.}
Contribution-level comparisons evaluate each specific claim extracted in Phase~I against retrieved candidates to assess novelty refutability. We adopt a one-to-N comparison approach: the target paper is compared against each candidate independently in separate inference calls. For each comparison, the LLM receives the normalized full texts of both papers (with references and acknowledgements removed) along with all contribution claims from Phase~I. This design isolates each comparison to prevent cross-contamination of judgments and enables parallel processing.

\paragraph{Judgment Categories.}
For each contribution, the LLM outputs a three-way judgment. \texttt{can\_refute} indicates that the candidate paper presents substantially similar ideas, methods, or findings that challenge the novelty claim. \texttt{cannot\_refute} indicates that the candidate paper, while related, does not provide sufficient evidence to challenge the claim. \texttt{unclear} indicates that the relationship cannot be determined due to insufficient information or ambiguous overlap.

\paragraph{Evidence Requirements.}
When the judgment is \texttt{can\_refute}, the LLM must provide explicit evidence in the form of verbatim quotes from both papers. Each evidence pair consists of a quote from the target paper (the claimed contribution) and a corresponding quote from the candidate paper (the prior work that challenges the claim). The LLM also provides a brief summary explaining the nature of the overlap.

\refstepcounter{examplebox}
\begin{tcolorbox}[
    colback=headerblue!10, 
    colframe=headerblue!50, 
    title={Example~\theexamplebox: Comparative Analysis Outputs},
    fonttitle=\bfseries,
    breakable,
    enhanced,
    float,                 
    floatplacement={!t},
]
\label{box:comparison_example}
\small
\textbf{A. Core-Task Distinction (Sibling-level):}
\begin{itemize}[leftmargin=*, nosep]
    \item \texttt{candidate\_paper\_title}: ``SkyRL-Agent: Efficient RL Training for Multi-turn...''
    \item \texttt{is\_duplicate\_variant}: \texttt{false}
    \item \texttt{brief\_comparison}: ``Both belong to the General-Purpose Multi-Turn RL Frameworks category. AgentGym-RL emphasizes the ScalingInter-RL method for horizon scaling, while SkyRL-Agent focuses on an optimized asynchronous pipeline dispatcher for efficiency gains.''
\end{itemize}

\vspace{0.5em}
\textbf{B. Contribution-level Comparison:} \\
\textit{Claim: ``AgentGym-RL framework for multi-turn RL-based agent training''}
\begin{itemize}[leftmargin=*, nosep]
    \item \texttt{candidate\_paper\_title}: ``Long-Horizon Interactive Agents''
    \item \texttt{refutation\_status}: \texttt{cannot\_refute}
    \item \texttt{brief\_note}: ``Long-Horizon Interactive Agents focuses on training interactive digital agents in a single stateful environment (AppWorld) using a specialized RL variant (LOOP). The original paper presents a broader multi-environment framework spanning web navigation, games, and embodied tasks.''
\end{itemize}
\end{tcolorbox}

\paragraph{Evidence Verification.}
LLM-generated quotes are verified against the source texts to prevent hallucinated citations. 
We employ a custom fuzzy matching algorithm based on token-level anchor alignment. 
A quote is considered verified if its confidence score exceeds 0.6, computed as a weighted 
combination of anchor coverage and hit ratio (details in Appendix~\ref{app:evidence_verification}). 
Only verified quotes appear in the final report; any \texttt{can\_refute} judgment lacking 
verified evidence is automatically downgraded to \texttt{cannot\_refute} at the end of Phase~III, 
before report generation.

\subsubsection{Implementation and Output}

Phase~III involves several technical considerations: \textbf{LLM-driven taxonomy construction} with MECE validation and automatic repair, \textbf{survey-style narrative synthesis} grounded in the taxonomy, \textbf{candidate-by-candidate comparisons} for both core-task positioning and contribution-level refutability, and \textbf{verification-first reporting} that checks LLM-produced evidence quotes and textual-overlap segments via a token-level alignment matcher (automatically downgrading \texttt{can\_refute} judgments when evidence cannot be verified). Appendix~\ref{app:phase3} details the corresponding specifications, including taxonomy node definitions, output schemas for contribution-level and core-task comparisons, evidence and similarity verification procedures, taxonomy-based comparison paths, prompt templates, and the complete Phase~III report schema.

The rationale for adopting \textbf{LLM-driven taxonomy} over traditional clustering, \textbf{three-way judgment classification}, \textbf{contribution-level focus}, \textbf{evidence verification as a hard constraint}, and \textbf{structured assembly} is discussed in Section~\ref{sec:discuss_analysis}; limitations regarding \textbf{mathematical formulas}, \textbf{visual content analysis}, and \textbf{taxonomy stability} are addressed in Section~\ref{sec:limit_content}.

Phase~III produces all analytical content: a hierarchical taxonomy with narrative synthesis, contribution-level comparisons with verified evidence, core-task distinctions, textual similarity analysis, and an overall novelty assessment. These results are serialized as a structured JSON file that serves as the sole input for \textbf{Phase~IV} rendering.

\subsection{Phase~IV: Report Generation}
\label{sec:phase4}

The fourth phase renders the Phase~III outputs into human-readable report formats. This phase involves \textbf{no LLM calls}; all analytical content has been finalized in Phase~III.

\subsubsection{Report Structure}

The final report renders seven modules from the Phase~III JSON: \texttt{original\_paper} (metadata), \texttt{core\_task\_survey} (taxonomy tree with 2-paragraph narrative), \texttt{contribution\_analysis} (per-contribution judgments with 3--4 paragraph overall assessment), \texttt{core\_task\_comparisons} (sibling paper distinctions), \texttt{textual\_similarity} (textual overlap analysis), \texttt{references} (unified citation index), and \texttt{metadata} (generation timestamps). All content is read directly from Phase~III outputs; no text is generated in this phase.

\subsubsection{Rendering}

The rendering process transforms the structured Phase~III output into human-readable formats. Key formatting decisions include: (1) consistent citation formatting across all modules (e.g., ``AgentGym-RL [1]''), (2) truncation of long evidence quotes for readability, and (3) hierarchical indentation for taxonomy visualization. The output can be rendered as Markdown or PDF. 

\subsubsection{Implementation and Output.}

Phase~IV is purely template-based: it reads the structured JSON produced by Phase~III and renders formatted reports using deterministic rules, without any LLM calls or analytical processing. Detailed rendering specifications are provided in Appendix~\ref{app:phase4}. 

Considerations regarding \textbf{pipeline dependencies} is addressed in Section~\ref{sec:limit_dependencies}.

The final output consists of Markdown and PDF reports. All reports are published on our website with full traceability, enabling users to verify any judgment by examining the cited evidence.

\section{Discussion}
\label{sec:discussion}

This section discusses design decisions, limitations, and ethical considerations of \textsc{OpenNovelty}.

\subsection{Design}
\label{sec:design}
This subsection explains the key design decisions made in each phase of \textsc{OpenNovelty}, including the rationale behind our choices and trade-offs considered.

\subsubsection{Extraction Design}
\label{sec:discuss_extraction}

\paragraph{Zero-shot Paradigm.}
We adopt a zero-shot extraction approach without few-shot examples. Few-shot examples may bias the model toward specific contribution types or phrasings present in the demonstrations, reducing generalization across diverse research areas. Zero-shot prompts are also easier to maintain and audit, as all behavioral specifications are contained in explicit natural language instructions rather than implicit patterns in examples.

\paragraph{Query Expansion Strategy.}
We generate semantic variants for each query rather than relying on a single precise query. This design reflects the inherent ambiguity in academic terminology, where the same concept may be expressed using different phrases across subfields. The primary query captures the exact terminology used in the target paper, while variants expand coverage to related phrasings that prior work may have used.

\subsubsection{Retrieval Design}
\label{sec:discuss_retrieval}

\paragraph{Broad Recall Strategy.}
We submit multiple queries per paper (typically 6--12 queries combining the core task and contributions with semantic variants), retrieving all candidates returned by the academic search API before applying multi-layer filtering. This design prioritizes recall over precision in the retrieval phase, delegating fine-grained relevance assessment to Phase~III. The rationale is that a missed relevant paper cannot be recovered in later phases, while irrelevant papers can be filtered out through subsequent analysis.

\paragraph{Semantic Relevance over Citation Metrics.}
We rely exclusively on semantic relevance scores rather than citation counts or venue prestige for candidate filtering. Citation counts are strongly time-dependent and systematically disadvantage recent papers. A highly relevant paper published one month ago may have zero citations, yet for novelty assessment, recent work is often more important than highly-cited older work. Similarly, venue filtering may exclude relevant cross-disciplinary work or important preprints.

\paragraph{Natural Language Queries.}
We use pure natural language queries rather than Boolean syntax (AND/OR/NOT). \textsc{Wispaper} is optimized for natural language input, where Boolean expressions may be parsed literally rather than interpreted semantically. Additionally, the semantic variants generated in Phase~I naturally achieve query expansion through paraphrasing and synonym substitution, eliminating the need for manual OR-based enumeration. Natural language queries are also more robust to phrasing variations and easier to maintain than brittle Boolean expressions.

\subsubsection{Analysis Design}
\label{sec:discuss_analysis}

\paragraph{LLM-driven Taxonomy.}
We adopt end-to-end LLM generation for taxonomy construction rather than traditional clustering algorithms such as K-means or hierarchical clustering. Taxonomy construction requires simultaneous consideration of methodology, research questions, and application domains, which embedding-based similarity cannot capture jointly. Additionally, hierarchical clustering produces distance-based splits without semantic labels, whereas LLMs generate meaningful category names and scope definitions for every node.

\paragraph{Three-way Judgment Classification.}
We use a three-way classification (\texttt{can\_refute}, \texttt{cannot\_refute}, \texttt{unclear}) rather than finer-grained similarity scores. Reviewers need clear, actionable signals about whether prior work challenges a novelty claim, not continuous scores that require interpretation thresholds. The \texttt{can\_refute} judgment requires explicit evidence quotes, ensuring that every positive finding is verifiable. Partial overlap scenarios are captured through \texttt{cannot\_refute} with explanatory summaries.

\paragraph{Contribution-level Focus.}
\label{sec:discuss_scope}
Unlike traditional surveys that compare papers along fixed axes such as architecture, accuracy, or computational complexity, \textsc{OpenNovelty} focuses exclusively on contribution-level refutability. Dimensions such as model architecture differences, experimental settings, and evaluation metrics are mentioned only when directly relevant to refuting a specific novelty claim. This design reflects our goal of novelty verification rather than comprehensive survey generation.

\paragraph{Evidence Verification as Hard Constraint.}
We treat evidence verification as a hard constraint rather than a soft quality signal. Any \texttt{can\_refute} judgment without verified evidence is automatically downgraded to \texttt{cannot\_refute} at the end of Phase~III, implemented as a final policy check after all LLM comparisons complete. This conservative approach may increase false negatives but ensures that every published refutation claim is verifiable. We prioritize precision over recall for refutation claims, as false positives are more harmful to authors than false negatives.

\paragraph{Structured Assembly.}
We generate report content through multiple independent LLM calls rather than a single end-to-end generation. This design enables targeted prompting for each component, independent retry and error handling, and parallel processing to reduce total generation time. Structured components such as statistics and citation indices are assembled programmatically from earlier phases.

\subsection{Limitations}
\label{sec:limitations}

Despite its capabilities, \textsc{OpenNovelty} has several limitations that users should be aware of when interpreting the generated reports.

\subsubsection{Content Analysis Constraints}
\label{sec:limit_content}

\paragraph{Mathematical Formulas.}
PDF text extraction produces unstable results for equations, symbols, subscripts, and matrices. Extracted formula fragments are processed as plain text without structural reconstruction. Consequently, the system cannot determine formula semantic equivalence or reliably detect novelty claims that hinge on mathematical innovations. This affects both Phase~I (contribution extraction) and Phase~III (claim comparison).

\paragraph{Visual Content.}

Figures, tables, architecture diagrams, and algorithm pseudocode are not analyzed. Contributions that are primarily communicated through visual representations may not be fully captured during extraction (Phase~I) or comparison (Phase~III). This limitation is particularly relevant for papers in computer vision, systems architecture, and algorithm design, where visual diagrams often convey core contributions. Users should interpret novelty assessments for visually-oriented papers with appropriate caution.

\paragraph{Taxonomy Stability.}

LLM-generated taxonomies may vary across runs due to the inherent stochasticity of language models. While MECE validation and automatic repair mechanisms mitigate structural issues (e.g., missing papers, duplicate assignments), category boundaries and hierarchical organization remain inherently subjective. Different runs may produce semantically valid but structurally different taxonomies, potentially affecting core-task comparison results. Taxonomies that fail validation after repair are flagged as \texttt{needs\_review} but still proceed through the pipeline with diagnostic annotations.

\subsubsection{Retrieval Boundaries}
\label{sec:limit_retrieval}

\paragraph{Indexing Scope.}
The system's novelty assessment is bounded by the semantic search engine's coverage. Papers not indexed by \textsc{Wispaper}, such as very recent preprints, non-English papers, or content in domain-specific repositories, cannot be retrieved or compared. This creates blind spots for certain research communities and publication venues.

\paragraph{Result Interpretation.}
A \texttt{cannot\_refute} judgment indicates that no retrieved paper refutes the claim, not that no such paper exists in the broader literature. Users should interpret this judgment as evidence that the system found no conflicting prior work within its search scope, rather than as definitive confirmation of novelty.

\subsubsection{Pipeline Dependencies}
\label{sec:limit_dependencies}

\paragraph{Error Propagation.}

The pipeline architecture introduces cascading dependencies where upstream errors affect downstream quality:
\begin{itemize}[nosep]
    \item \textbf{Phase~I $\to$ II}: Overly specific terminology or incomplete contribution claims can cause semantic search to miss relevant candidates.
    \item \textbf{Phase~II $\to$ III}: Aggressive quality filtering or temporal cutoff errors may exclude relevant prior work.
    \item \textbf{Within Phase~III}: Taxonomy misclassification can place the target paper among semantically distant siblings.
    \item \textbf{Phase~III $\to$ IV}: Phase~IV performs template-based rendering without additional analysis; errors from earlier phases propagate unchanged to the final report.
\end{itemize}

\paragraph{Mitigation Strategies.}

We partially mitigate these dependencies through query expansion (generating multiple semantic variants per item) and broad retrieval (Top-50 for core task, Top-10 per contribution). However, systematic extraction failures for certain paper types (e.g., unconventional structure, ambiguous contribution statements) cannot be fully addressed without manual intervention. The evidence verification step provides an additional safeguard by automatically downgrading unverified refutations (see Section~\ref{sec:discuss_analysis}).

\paragraph{Systemic Bias and Transparency.}
Beyond technical error propagation, the pipeline inherits systemic biases from its upstream components. The retrieval engine and LLM may exhibit preferences for well-indexed venues and English-language publications, creating blind spots in the novelty assessment. We address this by enforcing full traceability: all reports include relevance scores, taxonomy rationales, and evidence quotes. While this does not eliminate inherited bias, it ensures that the system's limitations are transparent to the user.

\subsection{Ethical Considerations}
\label{sec:ethics}
We discuss the ethical implications of deploying \textsc{OpenNovelty} in academic peer review.

\paragraph{Assistance vs. Replacement.}
\textsc{OpenNovelty} is intended as a retrieval and evidence-checking aid, not a decision-making system. It focuses on whether specific novelty claims can be challenged by retrieved prior work, but it does not measure broader research quality (e.g., significance, methodological rigor, clarity, or reproducibility). As a result, a \texttt{can\_refute} outcome does not imply that a paper lacks merit, and a \texttt{cannot\_refute} outcome does not establish novelty or importance. Final decisions should remain with human reviewers and area chairs.

\paragraph{Risks of Gaming and Over-reliance.}
We recognize two practical risks. First, authors could write in ways that weaken retrieval---for example, by using vague terminology or omitting key connections to prior work. Second, reviewers may exhibit automation bias and treat system outputs as authoritative. We reduce these risks through transparency: reports surface the retrieved candidates, relevance signals, and the specific evidence used to support each judgment, so that users can audit the basis of the conclusions.

\paragraph{Mandate for Constructive Use.}
To avoid harm, we explicitly discourage adversarial or punitive use of the reports. Because judgments can be affected by retrieval coverage and LLM interpretation errors, \textsc{OpenNovelty} should not be used as a standalone justification for desk rejection, nor as a tool to attack authors. Any \texttt{can\_refute} finding must be manually checked against the cited sources and interpreted in context. The goal is to support careful review by surfacing relevant literature, not to replace deliberation with automated policing.
\section{Future Work}
\label{sec:future_work}

While \textsc{OpenNovelty} provides a complete system for automated novelty analysis, several research directions remain open for systematic investigation.
 We outline our planned experiments organized by phase.

\subsection{Phase~I: Extraction Optimization}
The quality of downstream analyses fundamentally depends on accurate extraction of core tasks and contribution claims. We plan to conduct ablation studies across two dimensions.

\paragraph{Input Scope.}
We will compare extraction quality across varying input granularities, ranging from abstract-only to full text, with intermediate settings that include the introduction or conclusion. Our hypothesis is that abstracts provide sufficient signal for core task extraction, while contribution claims may benefit from broader textual context. Evaluation will measure extraction completeness, accuracy, and downstream retrieval effectiveness.

\paragraph{Extraction Strategies.}
We will evaluate alternative extraction approaches, contrasting zero-shot with few-shot prompting, single-pass with iterative refinement, and structured extraction with free-form generation followed by post-processing. The goal is to identify strategies that balance extraction quality with computational cost.

\subsection{Phase~II: Retrieval Benchmarking}

Current retrieval evaluation is limited by the absence of ground-truth relevance judgments. We plan to address this through benchmark construction and comparative analysis.

\paragraph{NoveltyBench Construction.}
We will construct \textsc{NoveltyBench}, a benchmark aggregating peer review data from two sources: (1) ML conferences via OpenReview, and (2) journals with transparent peer review policies (\textit{Nature Communications}, \textit{eLife}). The benchmark will include papers where reviewers cite specific prior work against novelty claims, cases where multiple reviewers independently identify the same related work, and author rebuttals acknowledging overlooked references. This enables evaluation of retrieval recall against human-identified relevant papers.

\paragraph{Search Engine Comparison.}
Using \textsc{NoveltyBench}, we will systematically compare retrieval effectiveness across multiple academic search engines, including Semantic Scholar, Google Scholar, OpenAlex, and \textsc{Wispaper}. Metrics will include recall@K, mean reciprocal rank, and coverage of reviewer-cited papers.

\subsection{Phase~III: Analysis Quality}
Phase~III encompasses multiple analytical components, each requiring dedicated evaluation.

\paragraph{Taxonomy Organization.}
We will investigate alternative taxonomy construction approaches, including iterative refinement with human feedback and hybrid methods that combine embedding-based clustering with LLM-generated labels. We will also explore different hierarchical depths and branching factors. Evaluation will assess both structural validity through MECE compliance checks and semantic coherence through human ratings.

\paragraph{Textual Similarity Detection.}
We will benchmark similarity detection accuracy against established plagiarism detection tools and manually annotated corpora. Key questions include optimal segment length thresholds, the trade-off between precision and recall, and effectiveness across verbatim versus paraphrased similarity types.

\paragraph{Refutation Judgment Calibration.}
We will evaluate the calibration of \texttt{can\_refute} judgments by comparing system outputs against expert assessments on a held-out set. This includes analyzing false positive and false negative rates, identifying systematic biases, and exploring confidence calibration techniques.

\subsection{End-to-End Evaluation}

Beyond component-level analysis, we plan comprehensive end-to-end evaluation:

\begin{itemize}
    \item \textbf{Agreement with Human Reviewers:} Measuring correlation between \textsc{OpenNovelty} assessments and reviewer novelty scores on OpenReview submissions.
    \item \textbf{User Studies:} Conducting studies with reviewers to assess whether \textsc{OpenNovelty} reports improve review efficiency and quality.
    \item \textbf{Longitudinal Analysis:} Tracking how novelty assessments correlate with eventual acceptance decisions and post-publication impact.
\end{itemize}

These experiments will inform iterative improvements to \textsc{OpenNovelty} and contribute benchmarks and insights to the broader research community.
\section{Related Work}
\label{sec:related work}

\subsection{AI-Assisted Peer Review}
The use of large language models in academic peer review has attracted increasing attention, with surveys showing that LLMs can produce review-like feedback across multiple dimensions \cite{zhuang2025large}. However, systematic evaluations reveal substantial limitations: LLM-as-a-Judge systems exhibit scoring bias that inflates ratings and distorts acceptance decisions \cite{li2025evaluating}, focus-level frameworks identify systematic blind spots where models miss nuanced methodological flaws \cite{shin2025mind}, and technical evaluations show particular difficulty identifying critical limitations requiring deep domain understanding \cite{xu2025can}. Large-scale monitoring studies document detectable stylistic homogenization in AI-modified reviews \cite{liang2024monitoring}, prompting policy responses from major venues \cite{iclr2026_llm_response, aaai2025ai_review}. While broader frameworks position LLMs as potential evaluators or collaborators in scientific workflows \cite{zhang2025evolving}, these limitations motivate approaches that restrict LLMs to well-defined, auditable subtasks. OpenNovelty follows this direction as an LLM-powered system for verifiable novelty assessment that grounds all judgments in retrieved real papers through claim-level full-text comparisons with traceable citations.

\subsection{Novelty Assessment}
Prior work can be organized into several complementary paradigms. (1) Detection and prediction approaches model novelty using statistical or learned signals \cite{liu2025harnessing}, with systematic studies showing that input scope and section combinations substantially affect assessment quality \cite{wu2025sc4anm}. (2) Retrieval-augmented methods leverage external evidence through two main strategies: comparative frameworks that formulate novelty as relative ranking between paper pairs using RAG over titles and abstracts \cite{lin2025evaluating}, and dimensional decomposition approaches that assess interpretable aspects (new problems, methods, applications) but face constraints from limited context windows and small retrieval sets \cite{shahid2025literature}. Closely related, idea-evaluation systems such as ScholarEval assess soundness and contribution by searching Semantic Scholar and conducting dimension-wise comparisons, but their contribution analysis is largely performed at the abstract level to enable broad coverage \cite{moussa2025scholareval}. (3) Review-oriented assistance systems support human reviewers by identifying missing related work through keyword-based or LLM-generated queries, though they typically lack claim-level verification \cite{afzal2025beyond}. (4) System- and framework-level approaches operationalize novelty and idea evaluation within structured representations or end-to-end reviewing pipelines. Graph-based frameworks employ explicit graph representations to organize ideas and research landscapes, enabling interactive exploration of novelty and relationships among contributions \cite{da2025graphmind, feng2025grapheval, gordetzki2025llm}. Complementarily, infrastructure platforms and reviewing frameworks demonstrate how LLM-based evaluation can be integrated into real-world peer review workflows with varying degrees of human oversight and standardization \cite{paperreview_ai, cspaper, yu2024automated}.

OpenNovelty addresses these limitations by grounding all assessments in retrieved real papers rather than LLM parametric knowledge. The system performs claim-level full-text comparisons and synthesizes structured reports where every judgment is explicitly linked to verifiable evidence with traceable citations.
\section{Conclusion}
\label{sec:conclusion}

We presented \textsc{OpenNovelty}, an LLM-powered system for transparent, evidence-based novelty analysis. By grounding all judgments in retrieved real papers with explicit citations and verified evidence, the system ensures that novelty assessments are traceable and verifiable. We have deployed \textsc{OpenNovelty} on 500+ ICLR 2026 submissions with all reports publicly available on our website. We plan to scale this analysis to over 2,000 submissions throughout the ICLR 2026 review cycle, providing comprehensive coverage of the conference. Future work will focus on systematic evaluation through \textsc{NoveltyBench} and optimization of each component. We hope this work contributes to fairer and more consistent peer review practices.

\bibliographystyle{unsrtnat}  
\bibliography{opennovelty} 

\clearpage

\beginappendix
\section{Phase~I: Detailed Specifications}
\label{app:phase1}

This appendix provides detailed specifications for Phase~I, including extraction field definitions, temperature configurations, prompt design principles, and output validation mechanisms.

\subsection{Phase~I Output Field Definitions}
\label{app:phase1_output_fields}

Phase~I produces three types of structured outputs: a core task description, contribution claims, and semantic queries. Tables~\ref{tab:app_core_task_fields}--\ref{tab:app_query_fields} define the fields and constraints for each output type.

\subsubsection{Core Task Output}
\label{app:core_task_fields}

Table~\ref{tab:app_core_task_fields} defines the structured fields for core task extraction.

\begin{table}[ht]
\centering
\caption{Structured fields for core task extraction}
\label{tab:app_core_task_fields}
\small
\begin{tabular}{llp{8cm}}
\toprule
\textbf{Field} & \textbf{Constraint} & \textbf{Description} \\
\midrule
\texttt{text} & 5--15 words & Single phrase describing the main problem or challenge; uses abstract field terminology rather than paper-specific model names \\
\texttt{query\_variants} & 3 queries & Semantic variants for query expansion; includes the original core task text plus 2 alternative phrasings \\
\bottomrule
\end{tabular}
\end{table}

\subsubsection{Contribution Output}

\label{app:contribution_fields}

Table~\ref{tab:app_contribution_fields} defines the structured fields for contribution output. These fields are generated across two steps: the first four fields are extracted during contribution extraction, while the query-related fields are synthesized in the subsequent query generation step.

\begin{table}[ht]
\centering
\caption{Structured fields for contribution output}
\label{tab:app_contribution_fields}
\small
\begin{tabular}{llp{7.5cm}}
\toprule
\textbf{Field} & \textbf{Constraint} & \textbf{Description} \\
\midrule
\multicolumn{3}{l}{\textit{Generated during contribution extraction:}} \\[0.3em]
\texttt{name} & $\leq$ 15 words & Concise noun phrase summarizing the contribution \\
\texttt{author\_claim\_text} & $\leq$ 40 words & Verbatim excerpt directly quoted from the paper \\
\texttt{description} & $\leq$ 60 words & One to two sentence paraphrase preserving key terminology \\
\texttt{source\_hint} & n/a & Location tag (e.g., ``Abstract'', ``Introduction \S 1'') \\
\midrule
\multicolumn{3}{l}{\textit{Generated during query generation:}} \\[0.3em]
\texttt{prior\_work\_query} & 5--25 words & Search query for retrieving related prior work; must begin with ``Find papers about''; soft limit 15 words, hard limit 25 words \\
\texttt{query\_variants} & 3 queries & Semantic variants including the original \texttt{prior\_work\_query} plus 2 alternative phrasings \\
\bottomrule
\end{tabular}
\end{table}

\subsubsection{Query Output Format}
\label{app:query_fields}

Table~\ref{tab:app_query_fields} defines the format specifications for generated queries.

\begin{table}[ht]
\centering
\caption{Query format specifications by source type}
\label{tab:app_query_fields}
\small
\begin{tabular}{lp{4cm}p{6cm}}
\toprule
\textbf{Source} & \textbf{Format} & \textbf{Constraint} \\
\midrule
Core task & Short phrase (no prefix) & 5--15 words; expressed directly without ``Find papers about'' prefix \\
Contribution & ``Find papers about [topic]'' & 5--25 words (soft limit: 15 words); must include search prefix \\
\midrule
Variants & Same as primary & 2 variants per primary query; use alternative terminology and standard abbreviations \\
\bottomrule
\end{tabular}
\end{table}

\subsection{Temperature and Query Configuration}
\label{app:phase1_config}

Table~\ref{tab:app_phase1_temp} summarizes the temperature settings and query statistics for Phase~I. Lower temperatures (0.0--0.1) ensure deterministic outputs for extraction tasks, while a slightly higher temperature (0.2) introduces controlled diversity for semantic variant generation. Table~\ref{tab:app_phase1_temp}(b) shows the query count breakdown: each paper generates 6--12 queries depending on the number of extracted contributions.

\begin{table}[ht]
\centering
\caption{Phase~I temperature settings and query statistics}
\label{tab:app_phase1_temp}
\small
\begin{minipage}{0.48\textwidth}
    \centering
    (a) Temperature settings\\[0.5em]
    \begin{tabular}{lc}
    \toprule
    \textbf{Task} & \textbf{Temp.} \\
    \midrule
    Contribution extraction & 0.0 \\
    Core task extraction & 0.1 \\
    Primary query generation & 0.0 \\
    Semantic variant generation & 0.2 \\
    \bottomrule
    \end{tabular}
\end{minipage}
\hfill
\begin{minipage}{0.48\textwidth}
    \centering
    (b) Query count per paper\\[0.5em]
    \begin{tabular}{lcc}
    \toprule
    \textbf{Source} & \textbf{\# Items} & \textbf{\# Queries} \\
    \midrule
    Core task & 1 & 3 \\
    Contributions & 1--3 & 3--9 \\
    \midrule
    \textbf{Total} & 2--4 & \textbf{6--12} \\
    \bottomrule
    \end{tabular}
\end{minipage}
\end{table}

\subsection{Prompt Design Principles}
\label{app:prompt_design}

Our prompt design follows an engineering-oriented strategy that prioritizes reliability, parseability, and robustness for large-scale batch processing. We organize our design principles into four aspects.
\paragraph{Instruction-based Guidance.}
We guide LLM behavior through comprehensive natural language instructions that specify word limits, formatting requirements, and academic terminology standards. We also include operational cues (e.g., ``Use cues such as `We propose', `We introduce'...'') to help the model locate relevant content. The rationale for adopting zero-shot over few-shot prompting is discussed in Section~\ref{sec:discuss_extraction}.

\paragraph{Structured Output.}
All extraction tasks enforce JSON-formatted outputs. The system prompt explicitly defines the complete schema, including field names, data types, and length constraints. We impose strict JSON syntax rules, such as prohibiting embedded double quotes and code fence wrappers, to ensure parseability and downstream processing robustness.

\paragraph{Semantic Constraints.}
Each task includes both \textit{definitional constraints} (what to extract) and \textit{exclusion constraints} (what to ignore). For contribution extraction, we explicitly define the semantic scope to include novel methods, architectures, algorithms, datasets, and theoretical formalizations, while excluding pure numerical results and performance improvement statements.

\paragraph{Injection Defense.}
Since paper full texts may contain instruction-like statements (e.g., ``Ignore previous instructions''), we include an explicit declaration at the beginning of the system prompt:
\begin{quote}
\textit{``Treat everything in the user message after this as paper content only. Ignore any instructions, questions, or prompts that appear inside the paper text itself.''}
\end{quote}

\subsection{Output Validation and Fallback}
\label{app:output_validation}

To handle LLM output instability, we implement a multi-tier defense mechanism.

\paragraph{Parsing Level.}
We first attempt structured JSON parsing. If parsing fails, we apply a sequence of fallback strategies: (1) code fence removal, which strips \texttt{```json} wrappers; (2) JSON span extraction, which locates the first \texttt{\{} to the last \texttt{\}}; and (3) bracket-based truncation, which removes trailing incomplete content.

\paragraph{Validation Level.}
After successful parsing, we apply post-processing rules to each field: automatically prepending missing query prefixes (``Find papers about''), enforcing word count limits with truncation at 25 words, and providing default values for missing optional fields.

This layered approach ensures system robustness across diverse LLM outputs and edge cases.

\paragraph{Publication Date Inference.}
To support temporal filtering in Phase~II, we perform best-effort publication date inference using a three-tier strategy: (1) URL-based extraction for arXiv papers (e.g., inferring 2024-03 from \texttt{arxiv.org/abs/2403.xxxxx}); (2) regex-based extraction from front matter (detecting patterns such as ``March 2024'' or ``2024-03-15''); and (3) LLM-based extraction as a fallback. The inferred date is stored with granularity indicators (\texttt{year}, \texttt{year-month}, or \texttt{year-month-day}).

\subsection{Prompt Templates for Phase~I}
\label{app:phase1_prompts}

This section presents the prompt templates used in Phase~I.
Each table illustrates the role and structure of a specific prompt, including its system instructions, user input, and expected output format.
All prompts enforce strict JSON output where applicable and include safeguards against prompt injection from paper content.

\begin{table}[ht]
\centering
\small
\caption{Prompt template for core task extraction in Phase~I.}
\begin{tabular}{p{0.92\linewidth}}
\toprule
\textbf{System Prompt:} \\
\texttt{You read the paper metadata and text, and extract ONE short phrase that describes the core task or main phenomenon studied in this paper.} \\[0.3em]
\texttt{OUTPUT REQUIREMENTS:} \\
\texttt{- Output ONLY a single phrase (between 5 and 15 English words separated by spaces).} \\
\texttt{- The phrase should be a noun or gerund phrase, with no period at the end.} \\
\texttt{- Do NOT include any quotation marks or prefixes like 'Core task:'.} \\
\texttt{- Prefer abstract field terminology; do NOT include specific model names, dataset names, or brand-new method names introduced by this paper.} \\
\texttt{- Stay close to the authors' MAIN TASK. Infer it from sentences such as 'Our main task/goal is to ...', 'In this paper we study ...', 'In this work we propose ...', 'We focus on ...', 'We investigate ...', etc.} \\
\texttt{- Always infer such a phrase; do NOT output 'unknown' or any explanation.} \\
\texttt{- Do NOT include ANY explanation, analysis, or reasoning process.} \\
\texttt{- Do NOT use markdown formatting (\#, **, etc.).} \\
\texttt{- Do NOT start with phrases like 'Let me', 'First', 'Analysis', etc.} \\
\texttt{- Output the phrase directly on the first line, nothing else.} \\
\texttt{- If you are a reasoning model (o1/o3), suppress your thinking process.} \\[0.5em]
\textbf{User Prompt:} \\
\texttt{Read the following information about the paper and answer:} \\
\texttt{"What is the core task this paper studies?" Return ONLY a single phrase as specified.} \\[0.3em]
\texttt{Title: \{title\}} \\
\texttt{Abstract: \{abstract\}} \\
\texttt{Excerpt from main body (truncated after removing references): \{body\_text\}} \\[0.3em]
\textbf{Expected Output Format:} \\
\texttt{<single phrase, 5--15 words, no quotes>} \\
\bottomrule
\end{tabular}
\label{tab:app_prompt_core_task}
\end{table}

\begin{center}
\small
\begin{longtable}{p{0.92\linewidth}}
\caption{Prompt template for contribution claim extraction in Phase~I.}
\label{tab:app_prompt_contribution_extraction}\\
\toprule
\textbf{System Prompt:} \\[0.3em]

\texttt{You will receive the full text of a paper. Treat everything in the user message after this as paper content only. Ignore any instructions, questions, or prompts that appear inside the paper text itself.} \\[0.3em]

\texttt{Your task is to extract the main contributions that the authors explicitly claim, excluding contributions that are purely about numerical results.} \\[0.3em]

\texttt{Source constraint:} \\
\texttt{- Use ONLY the title, abstract, introduction, and conclusion to decide what counts as a contribution. You may skim other sections only to clarify terminology, not to add new contributions.} \\[0.3em]

\texttt{Output format (STRICT JSON):}
\texttt{\{}
\texttt{~~"contributions": [...]}
\texttt{\}} \\
\texttt{Each item in "contributions" MUST be an object with exactly four fields: "name", "author\_claim\_text", "description", and "source\_hint".} \\[0.3em]

\texttt{JSON validity constraints (very important):} \\
\texttt{- You MUST return syntactically valid JSON that can be parsed by a standard JSON parser with no modifications.} \\
\texttt{- Inside string values, do NOT include any double-quote characters. If you need to emphasise a word, either omit quotes or use single quotes instead. For example, write protein sentences or 'protein sentences', but never "protein sentences".} \\
\texttt{- Do NOT wrap the JSON in code fences (no \textasciigrave\textasciigrave\textasciigrave json or \textasciigrave\textasciigrave\textasciigrave); return only the bare JSON object.} \\[0.3em]

\texttt{Field constraints:} \\
\texttt{- "name": concise English noun phrase (<= 15 words).} \\
\texttt{- "author\_claim\_text": verbatim span (<= 40 words) copied from the title, abstract, introduction, or conclusion. Do NOT paraphrase.} \\
\texttt{- "description": 1--2 English sentences (<= 60 words) paraphrasing the contribution without adding new facts; use the authors' key terminology when possible.} \\
\texttt{- "source\_hint": short location tag such as "Title", "Abstract", "Introduction \S 1", or "Conclusion paragraph 2".} \\[0.3em]

\texttt{Extraction guidelines:} \\
\texttt{- Exclude contributions that only report performance numbers, leaderboard improvements, or ablations with no conceptual message.} \\[0.3em]

\texttt{If the paper contains fewer than three such contributions, return only those that clearly exist. Do NOT invent contributions.} \\[0.3em]

\texttt{- Scan the title, abstract, introduction, and conclusion for the core contributions the authors claim.} \\
\texttt{- Definition of contribution: Treat as a contribution only deliberate non-trivial interventions that the authors introduce, such as: new methods, architectures, algorithms, training procedures, frameworks, tasks, benchmarks, datasets, objective functions, theoretical formalisms, or problem definitions that are presented as the authors' work.} \\
\texttt{- Use cues such as "Our contributions are", "We propose", "We introduce", "We develop", "We design", "We build", "We define", "We formalize", "We establish".} \\
\texttt{- Merge duplicate statements across sections; each entry must represent a unique contribution.} \\[0.3em]

\texttt{General rules:} \\
\texttt{- Output up to three contributions.} \\
\texttt{- Never hallucinate contributions that are not clearly claimed by the authors.} \\
\texttt{- Output raw, valid JSON only (no code fences, comments, or extra keys).} \\[0.5em]

\midrule
\textbf{User Prompt:} \\[0.3em]

\texttt{Extract up to three contributions claimed in this paper. Return "contributions" with items that satisfy the rules above.} \\[0.3em]

\texttt{Title:} \\
\texttt{\{title\}} \\[0.3em]

\texttt{Main body text (truncated and references removed when possible):} \\
\texttt{\{body\_text\}} \\[0.5em]

\midrule
\textbf{Expected Output Format (JSON):} \\[0.2em]
\texttt{\{"contributions": [\{"name": "...", "author\_claim\_text": "...", "description": "...", "source\_hint": "..."\}, ...]\}} \\
\bottomrule
\end{longtable}
\end{center}

\begin{table}[ht]
\centering
\small
\caption{Prompt template for primary query generation (prior work query) in Phase~I.}
\begin{tabular}{p{0.92\linewidth}}
\toprule
\textbf{System Prompt:} \\
\texttt{You generate prior-work search queries for claim-level novelty checking.} \\
\texttt{Each claim is provided with name, author\_claim\_text, and description. Produce ONE query per claim ID.} \\[0.3em]
\texttt{Output format:} \\
\texttt{- Return STRICT JSON:} \\
\texttt{~~\{"queries": [} \\
\texttt{~~~~\{"id": "...", "prior\_work\_query": "..."\},} \\
\texttt{~~~~\{"id": "...", "prior\_work\_query": "..."\},} \\
\texttt{~~~~...} \\
\texttt{~~]\}} \\
\texttt{- Do NOT include any extra keys, comments, or surrounding text.} \\[0.3em]
\texttt{ID mapping:} \\
\texttt{- Each input section beginning with "- [ID]" defines one claim.} \\
\texttt{- You MUST produce exactly one object in "queries" for each such ID.} \\
\texttt{- Copy the ID string exactly (without brackets) into the "id" field.} \\
\texttt{- Do NOT add, drop, or modify any IDs.} \\[0.3em]
\texttt{Requirements for prior\_work\_query:} \\
\texttt{- English only, single line per query, 5--15 words. YOU must never exceed the limit of 15 words.} \\
\texttt{- Each query MUST begin exactly with the phrase "Find papers about" followed by a space.} \\
\texttt{- Do not include proper nouns or brand-new method names that originate from this paper; restate the intervention using general technical terms.} \\
\texttt{- Preserve the claim's key task/intervention/insight terminology (including any distinctive words from the claim name) and the critical modifiers from author\_claim\_text/description. Do NOT replace them with vague substitutes unless absolutely necessary.} \\
\texttt{- If the claim asserts a comparative insight, keep both sides of the comparison in the query.} \\
\texttt{- Avoid filler phrases such as "in prior literature" or "related work".} \\
\texttt{- Do not add constraints or speculate beyond what the claim states.} \\
\texttt{- Do NOT wrap the JSON output in triple backticks; return raw JSON only.} \\[0.5em]
\textbf{User Prompt:} \\
\texttt{Generate one query per claim for the following claims:} \\
\texttt{- [contribution\_1]} \\
\texttt{~~name: \{name\}} \\
\texttt{~~author\_claim\_text: \{claim\_text\}} \\
\texttt{~~description: \{description\}} \\[0.3em]
\textbf{Expected Output Format (JSON):} \\
\texttt{\{"queries": [\{"id": "contribution\_1", "prior\_work\_query": "Find papers about ..."\}, ...]\}} \\
\bottomrule
\end{tabular}
\label{tab:app_prompt_primary_query}
\end{table}

\begin{table}[ht]
\centering
\small
\caption{Prompt template for semantic query variant generation in Phase~I.}
\begin{tabular}{p{0.95\linewidth}}
\toprule
\textbf{System Prompt:} \\
\texttt{You are an expert at rewriting academic search queries. Your job is to produce multiple short paraphrases that preserve EXACTLY the same meaning. Return JSON only, in the format: \{"variants": ["..."]\}. Requirements: generate 2–3 variants; EACH variant MUST begin exactly with 'Find papers about ' (note the trailing space); each variant must contain between 5 and 15 English words separated by spaces, you must never exceed the limit of 15 words; paraphrases must be academically equivalent to the original: preserve the task/object, paradigm, conditions, and modifiers; use only established field terminology and standard abbreviations (e.g., RL for reinforcement learning, long-term for long-horizon, multi-step for multi-turn); do not invent new synonyms, broaden or narrow scope, or alter constraints; do not include proper nouns or brand-new method names that originate from this paper; do not introduce new attributes (e.g., efficient, survey, benchmark); and nothing may repeat the original sentence verbatim. Do NOT wrap the JSON in triple backticks; respond with raw JSON.} \\[0.5em]
\textbf{User Prompt:} \\
\texttt{Original query:} \\
\texttt{\{primary\_query\}} \\[0.3em]
\texttt{Please provide 2–3 paraphrased variants.} \\[0.3em]
\textbf{Expected Output Format (JSON):} \\
\texttt{\{"variants": ["Find papers about <variant 1>", "Find papers about <variant 2>"]\}} \\
\bottomrule
\end{tabular}
\label{tab:app_prompt_query_variants}
\end{table}

\FloatBarrier

\section{Phase~II: Detailed Specifications}
\label{app:phase2}

This appendix provides detailed specifications for Phase~II, including filtering statistics, fault tolerance mechanisms, and quality flag generation.

\subsection{Filtering Statistics}
\label{app:filtering_stats}

Table~\ref{tab:app_filtering} summarizes typical filtering statistics observed across our evaluation set. The overall filtering rate is approximately 90--95\%, with quality flag filtering contributing the largest reduction.

\begin{table}[ht]
\centering
\caption{Typical filtering statistics in Phase~II (per paper)}
\label{tab:app_filtering}
\small
\begin{threeparttable}
\begin{tabular}{lcc}
\toprule
\textbf{Filtering Layer} & \textbf{Typical Count} & \textbf{Reduction} \\
\midrule
Raw retrieval & 800--1,000 & n/a \\
After quality flag filtering & 100--200 & $\sim$80\% \\
After deduplication & 50--150 & $\sim$20--50\% \\
After self-reference removal & 49--99 & $\sim$1\% \\
After temporal filtering & 40--80 & $\sim$20\% \\
Final Top-K selection & $\leq 50 + 10N$\tnote{*} & n/a \\
\bottomrule
\end{tabular}
\begin{tablenotes}
\footnotesize
\item[*] $N$ denotes the number of contributions, which is typically 1--3. Actual counts may be lower if fewer candidates survive filtering.
\end{tablenotes}
\end{threeparttable}
\end{table}

\subsection{Fault Tolerance Mechanisms}
\label{app:fault_tolerance}

We implement multi-layer fault tolerance to ensure robust retrieval under various failure conditions.

\paragraph{Automatic Retry.}
Each query incorporates retry logic with configurable backoff: up to 8 attempts per query (\texttt{PHASE2\_MAX\_QUERY\_ATTEMPTS}), with an initial delay of 5 seconds (\texttt{RETRY\_DELAY}). A separate global retry limit (\texttt{MAX\_RETRIES=180}) governs session-level retries for high-concurrency scenarios. This handles transient network failures and rate limiting without manual intervention.

\paragraph{Token Management.}
API authentication tokens are automatically refreshed before expiration. The system monitors token validity and preemptively renews credentials to prevent authentication failures during batch processing.

\paragraph{Graceful Degradation.}
If a query fails after all retry attempts, the system logs the failure and continues processing the remaining queries. Partial results are preserved, allowing downstream phases to proceed with available candidates rather than failing entirely.

\paragraph{Detailed Logging.}
All retrieval attempts, including failures, are logged with timestamps, query content, response codes, and error messages. This enables post-hoc debugging and identification of systematic issues, such as specific query patterns that consistently fail.

\subsection{Quality Flag Generation}
\label{app:quality_flag}

Before applying filtering layers, we locally compute quality flags (\texttt{perfect}, \texttt{partial}, \texttt{no}) for each retrieved paper based on the search engine's verification verdict. \textsc{Wispaper} returns a verdict containing a list of criteria assessments; each criterion has a \texttt{type} (e.g., \texttt{time}) and an assessment label such as \textit{support}, \textit{somewhat\_support}, \textit{reject}, or \textit{insufficient\_information}. We first apply a pre-filter using the ``verification-condition matching'' module; papers flagged as \texttt{no} are removed and do not participate in subsequent ranking.

A paper is marked as \texttt{perfect} if \emph{all} criteria assessments are \textit{support}. When there is only one criterion, the paper is marked as \texttt{partial} iff the assessment is \textit{somewhat\_support}; otherwise (i.e., \textit{reject} or \textit{insufficient\_information}) it is marked as \texttt{no}. When there are multiple criteria, the paper is marked as \texttt{partial} iff there exists at least one criterion with assessment in \{\textit{support}, \textit{somewhat\_support}\} whose \texttt{type} is not \texttt{time}; otherwise it is marked as \texttt{no}. This local computation ensures consistent quality assessment across all retrieved papers.

\FloatBarrier

\begin{table}[!htbp]
\centering
\small
\caption{Quality-flag mapping rules from \textsc{Wispaper}'s verification verdict.}
\label{tab:quality_flag_mapping}
\setlength{\tabcolsep}{0.7em}
\begin{tabular}{ll}
\toprule
\textbf{Condition} & \textbf{Quality Flag} \\
\midrule
All criteria assessments are \textit{support} & \texttt{perfect} \\
\midrule
\multicolumn{2}{l}{\textit{Single-criterion verdict (\#criteria = 1):}} \\
Assessment is \textit{somewhat\_support} & \texttt{partial} \\
Assessment is \textit{reject} or \textit{insufficient\_information} & \texttt{no} \\
\midrule
\multicolumn{2}{l}{\textit{Multi-criterion verdict (\#criteria $>$ 1):}} \\
$\exists$ a criterion with assessment in \{\textit{support}, \textit{somewhat\_support}\} and \texttt{type} $\neq$ \texttt{time} & \texttt{partial} \\
Otherwise & \texttt{no} \\
\bottomrule
\end{tabular}
\end{table}

\section{Phase~III: Detailed Specifications}
\label{app:phase3}

This appendix provides detailed specifications for Phase~III, including taxonomy node definitions, comparison output schema, evidence verification algorithms, and comparison path definitions.

\subsection{Taxonomy Node Definitions}
\label{app:taxonomy_nodes}

Table~\ref{tab:app_taxonomy_nodes} defines the structured fields for each node type in the taxonomy hierarchy.

\begin{table}[ht]
\centering
\caption{Field definitions for taxonomy nodes}
\label{tab:app_taxonomy_nodes}
\small
\setlength{\tabcolsep}{0.6em}
\begin{tabular}{llp{6.1cm}}
\toprule
\textbf{Node Type} & \textbf{Field} & \textbf{Description} \\
\midrule
Root & \texttt{name} & ``\{TOPIC\_LABEL\} Survey Taxonomy'' format \\
& \texttt{subtopics} & List of child branch nodes \\
\midrule
\multirow{4}{*}{Internal} & \texttt{name} & Category name (3--8 words) \\
& \texttt{scope\_note} & Inclusion rule ($\leq$ 25 words) \\
& \texttt{exclude\_note} & Exclusion rule with redirect ($\leq$ 25 words) \\
& \texttt{subtopics} & List of child nodes (non-empty) \\
\midrule
\multirow{4}{*}{Leaf} & \texttt{name} & Cluster name (3--8 words) \\
& \texttt{scope\_note} & Inclusion rule ($\leq$ 25 words) \\
& \texttt{exclude\_note} & Exclusion rule with redirect ($\leq$ 25 words) \\
& \texttt{papers} & List of 2--7 paper identifiers \\
\bottomrule
\end{tabular}
\end{table}

\subsection{Claimed Contribution Comparison Output Schema}
\label{app:comparison_schema}

Table~\ref{tab:app_comparison_schema} defines the structured output format for each pairwise claimed contribution comparison.

\begin{table}[ht]
\centering
\caption{Output schema for contribution-level comparison (per candidate)}
\label{tab:app_comparison_schema}
\small
\setlength{\tabcolsep}{0.6em}
\begin{tabular}{ll}
\toprule
\textbf{Field} & \textbf{Description} \\
\midrule
\texttt{canonical\_id} & Canonical identifier of the candidate paper \\
\texttt{candidate\_paper\_title} & Title of the candidate paper \\
\texttt{candidate\_paper\_url} & URL to the candidate paper \\
\texttt{comparison\_mode} & Mode used: \texttt{fulltext} or \texttt{abstract} \\
\texttt{refutation\_status} & One of: \texttt{can\_refute}, \texttt{cannot\_refute}, \texttt{unclear} \\
\texttt{refutation\_evidence} & Object with \texttt{summary} and \texttt{evidence\_pairs} (if \texttt{can\_refute}) \\
\texttt{brief\_note} & 1--2 sentence explanation (if not \texttt{can\_refute}) \\
\texttt{similarity\_segments} & List of detected textual overlap segments (merged from detection module) \\
\bottomrule
\end{tabular}
\end{table}

\subsection{Core-Task Comparison Output Schema}
\label{app:core_task_schema}

Table~\ref{tab:app_core_task_schema} defines the structured output format for core-task comparisons, which differ from contribution-level comparisons in structure and purpose.

\begin{table}[ht]
\centering
\caption{Output schema for core-task comparison (per sibling)}
\label{tab:app_core_task_schema}
\small
\setlength{\tabcolsep}{0.6em}
\begin{tabular}{ll}
\toprule
\textbf{Field} & \textbf{Description} \\
\midrule
\texttt{canonical\_id} & Canonical identifier of the sibling paper \\
\texttt{candidate\_paper\_title} & Title of the sibling paper \\
\texttt{candidate\_paper\_url} & URL to the sibling paper \\
\texttt{relationship} & Relationship type (\texttt{sibling}) \\
\texttt{comparison\_mode} & Mode used: \texttt{fulltext} or \texttt{abstract\_fallback} \\
\texttt{is\_duplicate\_variant} & Boolean: \texttt{true} if papers appear to be versions of the same work \\
\texttt{brief\_comparison} & 2--3 sentence summary of methodological/conceptual distinctions \\
\texttt{similarity\_segments} & List of detected textual overlap segments (merged from detection module) \\
\bottomrule
\end{tabular}
\end{table}

\subsection{Evidence Verification Algorithm}
\label{app:evidence_verification}

We verify LLM-generated quotes using a fuzzy matching algorithm based on token-level sequence alignment.

\paragraph{Tokenization.}
Both the generated quote and the source document are tokenized using whitespace and punctuation boundaries. We normalize tokens by converting to lowercase and removing leading and trailing punctuation.

\paragraph{Alignment.}
We employ the SequenceMatcher algorithm from Python's difflib library to find the longest contiguous matching subsequence. The algorithm computes a similarity ratio defined as $2M / T$, where $M$ is the number of matching tokens and $T$ is the total number of tokens in both sequences.

\paragraph{Verification Criteria.}
A quote is considered verified if the confidence score exceeds 0.6. The confidence score is computed as:
\begin{equation}
\text{confidence} = \begin{cases}
0.7 \times \bar{c} + 0.3 \times h & \text{if anchors are compact} \\
0.5 \times (0.7 \times \bar{c} + 0.3 \times h) & \text{otherwise}
\end{cases}
\end{equation}
where $\bar{c}$ is the mean coverage of anchors achieving $\geq 60\%$ token-level match, $h$ is the fraction of anchors achieving such matches (hit ratio), and ``compact'' means all matched anchor spans have gaps $\leq 300$ tokens between consecutive anchors. The minimum anchor length is 20 characters. Quotes that fail verification are flagged with \texttt{found=false} in the location object; any \texttt{can\_refute} judgment lacking verified evidence is automatically downgraded to \texttt{cannot\_refute}.

\subsection{Similarity Verification Details}
\label{app:similarity_scoring}

\paragraph{Segment Verification.}

Each LLM-reported similarity segment undergoes verification using the same anchor-based algorithm described in Appendix~\ref{app:evidence_verification}. Both original and candidate quotes are normalized (lowercased, whitespace-collapsed, Unicode variants replaced) and verified independently against their respective source documents.

\paragraph{Acceptance Criteria.}

A segment is accepted only if it meets all of the following conditions:
\begin{itemize}
    \item Both quotes achieve at least 60\% token-level coverage in their respective source documents
    \item The minimum word count across both quotes is at least 30 words
    \item Both \texttt{found} flags in the location objects are \texttt{true}
\end{itemize}

\paragraph{Segment Filtering.}

If more than 3 segments pass verification, only the top 3 by word count are retained as representative evidence. Segments are classified as either ``Direct'' (verbatim copying) or ``Paraphrase'' (rewording without attribution) based on LLM assessment.

\subsection{Core-Task Comparison Path Definitions}

\label{app:comparison_paths}

The core-task comparison adapts its granularity based on the paper's structural position in the taxonomy, ensuring relevant context for distinction analysis. Table~\ref{tab:app_comparison_paths} defines these execution paths.

\begin{table}[ht]
\centering
\caption{Taxonomy-based comparison execution paths}
\label{tab:app_comparison_paths}
\small
\begin{tabular}{llp{6.4cm}}
\toprule
\textbf{Position Type} & \textbf{Condition} & \textbf{Analysis Level} \\
\midrule
\texttt{sibling}\textsuperscript{*} & Siblings in leaf $> 0$ & Individual paper-level distinction analysis against each sibling in the same leaf. \\
\texttt{subtopic\_siblings}\textsuperscript{*} & Siblings $= 0$, Subtopics $> 0$ & Categorical analysis between the target leaf and sibling subtopics under the same parent. \\
\texttt{isolated}\textsuperscript{*} & Both $= 0$ & No comparison; log structural isolation as the paper has no immediate semantic neighbors. \\
\bottomrule
\end{tabular}

\textsuperscript{*} Implementation uses: \texttt{has\_siblings}, \texttt{no\_siblings\_but\_subtopic\_siblings}, \texttt{no\_siblings\_no\_subtopic\_siblings}.
\end{table}

\subsection{Prompt Templates for Phase~III}
\label{app:phase3_prompts}

This section presents the prompt templates used in Phase~III.
Each table illustrates the role and structure of a specific prompt, including system instructions, user inputs, and expected output formats.
All prompts enforce strict JSON output and include MECE validation, evidence verification, and hallucination prevention constraints.

{\small
\setlength{\LTpre}{0pt}
\setlength{\LTpost}{0pt}
\begin{longtable}{p{0.95\textwidth}}
\caption{Prompt template for hierarchical taxonomy construction in Phase~III.}
\label{tab:app_prompt_taxonomy_construction}\\
\toprule
\textbf{System Prompt:} \\
\midrule
\endfirsthead
\caption[]{Prompt template for hierarchical taxonomy construction in Phase~III. (continued)}\\
\toprule
\textbf{System Prompt:} \\
\midrule
\endhead
\bottomrule
\endfoot
\bottomrule
\endlastfoot
\texttt{You are a senior survey researcher specializing in building rigorous academic taxonomies.} \\
\texttt{Return EXACTLY ONE JSON object and NOTHING ELSE (no markdown, no code fences, no explanations).} \\[0.3em]
\texttt{INPUT (user JSON)} \\
\texttt{- topic: a short core-task phrase (typically one line). Use it as the base topic label.} \\
\texttt{- original\_paper\_id: optional (may be null).} \\
\texttt{- papers: list of \{id, title, abstract, rank\}. IMPORTANT: ids are canonical identifiers; copy them verbatim.} \\[0.3em]
\texttt{HARD CONSTRAINTS (must satisfy all)} \\
\texttt{1) Output must be valid JSON parseable by json.loads: use DOUBLE QUOTES for all keys and all string values.} \\
\texttt{2) Output must follow the schema exactly. Do not add any extra keys.} \\
\texttt{3) Use ONLY the provided paper ids. Do NOT invent ids. Do NOT drop ids. Do NOT duplicate ids.} \\
\texttt{~~~Every unique input id must appear EXACTLY ONCE across ALL leaves.} \\
\texttt{4) Tree structure:} \\
\texttt{~~~- Root MUST have: name, subtopics.} \\
\texttt{~~~- Root MUST NOT contain: scope\_note, exclude\_note, papers.} \\
\texttt{~~~- Non-leaf nodes MUST have: name, scope\_note, exclude\_note, subtopics (non-empty).} \\
\texttt{~~~- Leaf nodes MUST have: name, scope\_note, exclude\_note, papers (non-empty array of ids).} \\
\texttt{~~~- Non-leaf nodes MUST NOT contain 'papers'. Leaf nodes MUST NOT contain 'subtopics'.} \\
\texttt{5) Root naming:} \\
\texttt{~~~- Set TOPIC\_LABEL by taking user.topic and, only if needed, compressing it to <= 8 words WITHOUT changing meaning.} \\
\texttt{~~~- Allowed compression operations: delete redundant modifiers/articles/auxiliary prepositional phrases.} \\
\texttt{~~~~~Do NOT replace core technical nouns/verbs with synonyms, and do NOT introduce new terms.} \\
\texttt{~~~- Root name MUST be exactly: "TOPIC\_LABEL Survey Taxonomy".} \\
\texttt{~~~- Do NOT output the literal string "<topic>" and do NOT include angle brackets.} \\
\texttt{6) If original\_paper\_id is provided, it must appear in exactly one leaf.} \\[0.3em]
\texttt{ACADEMIC BOTTOM-UP PROCEDURE (survey style)} \\
\texttt{A) Read titles+abstracts and extract key attributes per paper (use domain-appropriate equivalents):} \\
\texttt{~~~core approach / intervention type / theoretical framework; research question / objective / outcome or endpoint;} \\
\texttt{~~~evidence basis \& study design (experimental/empirical/theoretical; validation protocol/setting);} \\
\texttt{~~~data/materials/subjects and context (modality if applicable; application domain only if it separates papers meaningfully).} \\
\texttt{B) Create micro-clusters of highly similar papers.} \\
\texttt{C) Name each micro-cluster with precise technical terminology.} \\
\texttt{D) Iteratively abstract upward into parents, maintaining clear boundaries.} \\[0.3em]
\texttt{DEFAULT ORGANIZATION PRINCIPLE (domain-agnostic; override if corpus suggests otherwise)} \\
\texttt{Choose the top-level split by the axis that maximizes discriminability and interpretability for this corpus:} \\
\texttt{clear boundaries, reasonably balanced coverage, and minimal cross-membership.} \\
\texttt{Consider axes in this order of preference ONLY IF they yield strong boundaries:} \\
\texttt{1) Core approach / intervention type / theoretical framework (what is being proposed, changed, or assumed)} \\
\texttt{2) Research question / objective / outcome (what is being solved, measured, or optimized)} \\
\texttt{3) Study context and evidence basis (data/materials/subjects; experimental/empirical design; evaluation/validation protocol; setting)} \\
\texttt{If multiple axes tie, prefer the one that yields more stable, reusable categories across papers.} \\[0.3em]
\texttt{MECE REQUIREMENT AT EVERY SIBLING GROUP} \\
\texttt{- Mutually exclusive scopes; collectively exhaustive coverage under the parent.} \\
\texttt{- Each node must include:} \\
\texttt{~~- scope\_note: exactly ONE sentence (<= 25 words) stating a clear inclusion rule.} \\
\texttt{~~- exclude\_note: exactly ONE sentence (<= 25 words) stating a clear exclusion rule AND where excluded items belong.} \\[0.3em]
\texttt{NAMING RULES (important)} \\
\texttt{- Use concrete technical terms. Avoid vague buckets.} \\
\texttt{- Forbidden words in category names: other, others, misc, miscellaneous, general, uncategorized, unclear.} \\
\texttt{- Keep names concise (<= 5–7 words typically) but prioritize clarity.} \\[0.3em]
\texttt{STRUCTURE BALANCE (soft constraints; never override HARD constraints)} \\
\texttt{- Prefer depth 3–5.} \\
\texttt{- Typical leaf size 2–7. If a leaf would exceed \textasciitilde7 papers, split it using the next most informative axis.} \\
\texttt{- Leaf size 1 is allowed ONLY if the paper is semantically distinct and merging would blur boundaries; otherwise merge into the closest sibling.} \\[0.3em]
\texttt{NEAR-DUPLICATES / VERSIONS} \\
\texttt{- Different ids may be versions of the same work. KEEP all ids separate (no merging/deletion).} \\
\texttt{- You MAY place suspected versions under the same leaf if it is the best-fit category.} \\[0.3em]
\texttt{ORDERING WITHIN EACH LEAF'S 'papers'} \\
\texttt{- If original\_paper\_id is in that leaf, put it first.} \\
\texttt{- Then order remaining ids by ascending input rank (ties: preserve the original input order).} \\[0.3em]
\texttt{FINAL SELF-CHECK (mandatory before returning)} \\
\texttt{- The union of all leaf 'papers' arrays equals the set of unique input ids (exact set equality).} \\
\texttt{- No id appears twice; no unknown id appears; no empty subtopics; no empty papers.} \\
\texttt{- Root name matches exactly "TOPIC\_LABEL Survey Taxonomy".} \\[0.3em]
\texttt{OUTPUT SCHEMA (STRICT; valid JSON; no extra keys)} \\
\texttt{\{"name": "TOPIC\_LABEL Survey Taxonomy", "subtopics": [\{"name": "Parent", "scope\_note": "...", "exclude\_note": "...", "subtopics": [\{"name": "Child", "scope\_note": "...", "exclude\_note": "...", "subtopics": [\{"name": "Leaf", "scope\_note": "...", "exclude\_note": "...", "papers": ["<paper\_id>", "..."]\}]\}]\}]\}} \\[0.5em]
\textbf{User Prompt (JSON):} \\
\texttt{\{"topic": "<core\_task>", "original\_paper\_id": "<id or null>", "papers": [\{"id": "...", "title": "...", "abstract": "...", "rank": N\}, ...]\}} \\[0.3em]
\textbf{Expected Output Format (JSON):} \\
\texttt{\{"name": "TOPIC\_LABEL Survey Taxonomy", "subtopics": [\{"name": "...", "scope\_note": "...", "exclude\_note": "...", "subtopics": [...] | "papers": ["id1", "id2"]\}, ...]\}} \\
\end{longtable}
}

\begin{center}
\small
\begin{longtable}{p{0.95\linewidth}}
\caption{Prompt template for taxonomy repair in Phase~III.}
\label{tab:app_prompt_taxonomy_repair}\\
\toprule
\textbf{System Prompt:} \\
\texttt{You will receive a taxonomy JSON and constraints for fixing it.} \\
\texttt{Return EXACTLY ONE valid JSON object parseable by json.loads (DOUBLE QUOTES; no code fences; no extra text).} \\[0.3em]
\texttt{Hard constraints:} \\
\texttt{- Single root. Root name MUST exactly equal root\_name.} \\
\texttt{- Only leaf nodes may have 'papers' (non-empty array of ids). Non-leaf nodes must NOT have 'papers'.} \\
\texttt{- Internal nodes must have non-empty 'subtopics'. Leaf nodes must NOT have 'subtopics'.} \\
\texttt{- Use ONLY allowed\_ids. Remove any extra\_ids. Every allowed id must appear EXACTLY ONCE across all leaves.} \\
\texttt{- If original\_paper\_id is present in the taxonomy input, keep it assigned to exactly one leaf.} \\[0.3em]
\texttt{Repair style (important):} \\
\texttt{- MINIMAL-CHANGE: keep the existing node names and hierarchy whenever possible.} \\
\texttt{- First ensure extra\_ids are removed and duplicates are eliminated.} \\
\texttt{- Then place missing\_papers into the best-fit existing leaves using their titles/abstracts.} \\
\texttt{- Only if no existing leaf fits, create a small new leaf or a minimal new branch.} \\[0.5em]
\textbf{User Prompt (JSON):} \\[0.3em]
\texttt{\{"root\_name": "<taxonomy\_root\_name>", "allowed\_ids": ["id1", "id2", ...],} \\
\texttt{~"missing\_ids": ["id3", ...], "extra\_ids": ["id4", ...],} \\
\texttt{~"missing\_papers": [\{"id": "...", "title": "...", "abstract": "...", "rank": N\}, ...],} \\
\texttt{~"taxonomy": <original\_taxonomy\_json>\}} \\[0.3em]
\textbf{Expected Output Format (JSON):} \\[0.2em]
\texttt{\{"name": "TOPIC\_LABEL Survey Taxonomy", "subtopics": [...]\}} \\
\bottomrule
\end{longtable}
\end{center}

\begin{center}
\small
\begin{longtable}{p{0.95\linewidth}}
\caption{Prompt template for survey narrative synthesis in Phase~III.}
\label{tab:app_prompt_narrative_synthesis}\\
\toprule
\textbf{System Prompt:} \\
\texttt{You are writing a SHORT survey-style narrative for domain experts.} \\
\texttt{You will receive:} \\
\texttt{- A core task description (if available).} \\
\texttt{- A taxonomy root name and a list of top-level branches.} \\
\texttt{- The taxonomy path and leaf neighbors of the original paper.} \\
\texttt{- A citation\_index mapping each paper\_id to \{alias,index,year,is\_original\}.} \\[0.3em]
\texttt{Your job is to return STRICT JSON only (no code fences) with a single key:} \\
\texttt{\{} \\
\texttt{~~'narrative': '<TWO short paragraphs, plain text>'} \\
\texttt{\}} \\[0.3em]
\texttt{NARRATIVE REQUIREMENTS:} \\
\texttt{- Write EXACTLY TWO paragraphs, separated by a single blank line.} \\
\texttt{- No headings, no bullet points; use compact, fluent prose.} \\
\texttt{- Overall length should be relatively short (roughly 180--250 words).} \\
\texttt{- In the FIRST paragraph:} \\
\texttt{~~* Briefly restate the core task in your own words (you may start with} \\
\texttt{~~~~'Core task: <core\_task\_text>' when a core\_task\_text is provided).} \\
\texttt{~~* Give a high-level picture of the field structure suggested by the} \\
\texttt{~~~~taxonomy: what the main branches are, what kinds of methods or} \\
\texttt{~~~~problem settings each branch tends to focus on, and how they relate.} \\
\texttt{~~* You may mention a few representative works using the Alias[index]} \\
\texttt{~~~~style when it helps to make the structure concrete.} \\
\texttt{- In the SECOND paragraph:} \\
\texttt{~~* Zoom in on a few particularly active or contrasting lines of work,} \\
\texttt{~~~~and describe the main themes, trade-offs, or open questions that} \\
\texttt{~~~~appear across these branches.} \\
\texttt{~~* Naturally situate the original paper (typically Alias[0]) within this} \\
\texttt{~~~~landscape: describe which branch or small cluster of works it feels} \\
\texttt{~~~~closest to, and how its emphasis compares to one or two nearby} \\
\texttt{~~~~papers (for example Alias[3], Alias[5]), without rewriting a full} \\
\texttt{~~~~taxonomy path.} \\
\texttt{~~* Keep the tone descriptive: you are helping the reader see where this} \\
\texttt{~~~~work roughly sits among existing directions, not writing a review} \\
\texttt{~~~~decision.} \\[0.3em]
\texttt{NUMERIC STYLE:} \\
\texttt{- Avoid detailed integer counts of papers or branches; instead prefer} \\
\texttt{~~qualitative phrases such as 'a small handful of works', 'many studies',} \\
\texttt{~~'a dense branch', 'only a few papers'.} \\
\texttt{- You may use numeric indices that are part of citations like Alias[0]} \\
\texttt{~~or Alias[3]; this is allowed.} \\[0.3em]
\texttt{CITATION STYLE:} \\
\texttt{- When you want to mention a specific paper, use its Alias[index] from} \\
\texttt{~~citation\_index. For example: 'AgentGym-RL[0] provides ...',} \\
\texttt{~~'Webagent-r1[1] focuses on ...'.} \\
\texttt{- Do not invent new aliases; only use the provided ones.} \\
\texttt{- You may ONLY cite indices that are listed in allowed\_citation\_indices.} \\
\texttt{~~Do not cite any other index.} \\[0.5em]
\textbf{User Prompt (JSON):} \\
\texttt{\{"language": "en", "core\_task\_text": "...", "taxonomy\_root": "...", "top\_level\_branches": [...],} \\
\texttt{~"original\_paper\_id": "...", "original\_taxonomy\_path": [...], "neighbor\_ids": [...],} \\
\texttt{~"citation\_index": \{...\}, "allowed\_citation\_indices": [...]\}} \\[0.3em]
\textbf{Expected Output Format (JSON):} \\
\texttt{\{"narrative": "<Two paragraphs of survey-style prose>"\}} \\
\bottomrule
\end{longtable}
\end{center}

\begin{center}
\small
\begin{longtable}{p{0.95\linewidth}}
\caption{Prompt template for one-liner generation in Phase~III.}
\label{tab:app_prompt_one_liner_generation}\\
\toprule
\textbf{System Prompt:} \\
\texttt{Write a concise one-liner summary for each paper (20--30 words).} \\
\texttt{Return STRICT JSON only: \{"items": [\{"paper\_id": id, "brief\_one\_liner": text\}, ...]\}} \\
\texttt{with the SAME order and length as the input list.} \\
\texttt{Do not invent numbers; base only on title/abstract. Language: English.} \\[0.3em]
\texttt{JSON FORMAT RULES (CRITICAL):} \\
\texttt{- The entire response must be valid JSON that can be parsed by a standard json.loads implementation.} \\
\texttt{- Do NOT wrap the JSON in code fences such as or ```; return raw JSON only.} \\
\texttt{- Inside JSON string values, do not use unescaped double quotes.} \\
\texttt{~~~If you need quotes inside a string, either use single quotes or escape double quotes as \textbackslash".} \\
\texttt{- Do not include comments, trailing commas, or any keys beyond 'items', 'paper\_id', 'brief\_one\_liner'.} \\[0.5em]
\midrule
\textbf{User Prompt (JSON):} \\
\texttt{\{"papers": [\{"canonical\_id": "...", "title": "...", "abstract": "..."\}, ...]\}} \\[0.3em]
\textbf{Expected Output Format (JSON):} \\
\texttt{\{"items": [\{"paper\_id": "...", "brief\_one\_liner": "20--30 word summary..."\}, ...]\}} \\
\bottomrule
\end{longtable}
\end{center}

\begin{center}
\small
\begin{longtable}{p{0.95\linewidth}}
\caption{Prompt template for textual similarity (plagiarism) detection in Phase~III.}
\label{tab:app_prompt_similarity_detection}\\
\toprule
\textbf{Unified Prompt (Single User Message):} \\[0.3em]
\texttt{\# Role} \\[0.2em]
\texttt{You are an extremely rigorous academic plagiarism detection system, focused on **verbatim, character-level** text comparison. Your core capability is to precisely identify plagiarism segments between two input texts and extract them **without any alteration**.} \\[0.3em]
\texttt{\# Task} \\[0.2em]
\texttt{Compare the contents of **[Paper A]** and **[Paper B]**, and identify all paragraphs where the **continuous text overlap is very high**.} \\[0.3em]
\texttt{\# Input Format} \\[0.2em]
\texttt{The user will provide two text sections, labeled as `<Paper\_A>` and `<Paper\_B>`.} \\[0.3em]
\texttt{\# Strict Constraints (must be strictly followed)} \\[0.2em]
\texttt{1. **Absolutely no rewriting**: The `original\_text` and `candidate\_text` fields in the output must be **exactly identical** to the input texts, including punctuation and whitespace. It is strictly forbidden to summarize, polish, reorder, or replace words with synonyms.} \\[0.2em]
\texttt{2. **No hallucination**: Before outputting any segment, you must perform an internal "Ctrl+F-like" search. Every extracted segment must occur verbatim in the provided inputs. If the segment does not exist, you must not output it under any circumstances.} \\[0.2em]
\texttt{3. **Length threshold**: Only report segments where the number of **consecutive words is \textgreater= 30**. Ignore short phrases or coincidental overlaps of common terms.} \\[0.2em]
\texttt{4. **Plagiarism gate: Report a segment only when it meets both} \\
\texttt{~~~(i) **Verbatim overlap**: \textgreater= 30 consecutive-word verbatim overlap (after formula normalization), the unique wording pattern is substantially the same as the source.} \\
\texttt{~~~(ii) **Strict semantic/structural equivalence**: same claim, same technical entities, and same logical direction.} \\[0.2em]
\texttt{5. **Prefer omission over error**: If no overlapping segments that meet the criteria are found, return an empty list directly. Do not fabricate or force results.} \\[0.3em]
\texttt{\# Definition of Plagiarism} \\[0.2em]
\texttt{- **Direct Plagiarism**: Copying text word-for-word from a source without quotation marks and citation.} \\[0.2em]
\texttt{- **Paraphrasing Plagiarism**: Rewording a source's content without changing its core meaning or providing attribution.} \\[0.3em]
\texttt{\# Output Format} \\[0.2em]
\texttt{Output only a valid JSON object, with no Markdown code block markers (such as), and no opening or closing remarks.} \\[0.2em]
\texttt{The JSON structure is as follows:} \\[0.2em]
\texttt{**When plagiarism segments are found:**} \\
\texttt{\{} \\
\texttt{~~"plagiarism\_segments": [} \\
\texttt{~~~~\{} \\
\texttt{~~~~~~"segment\_id": 1,} \\
\texttt{~~~~~~"location": "Use an explicit section heading if it appears in the provided text near the segment (e.g., Abstract/Introduction/Method/Experiments). If no explicit heading is present, output \textbackslash"unknown\textbackslash".",} \\
\texttt{~~~~~~"original\_text": "This must be an exact quote that truly exists in Paper A, no modifications allowed...",} \\
\texttt{~~~~~~"candidate\_text": "This must be an exact quote that truly exists in Paper B, no modifications allowed...",} \\
\texttt{~~~~~~"plagiarism\_type": "Direct/Paraphrase",} \\
\texttt{~~~~~~"rationale": "Brief explanation (1-2 sentences) of why these texts are plagiarism."} \\
\texttt{~~~~\}} \\
\texttt{~~]} \\
\texttt{\}} \\[0.2em]
\texttt{**When NO plagiarism is detected:**} \\
\texttt{\{} \\
\texttt{~~"plagiarism\_segments": []} \\
\texttt{\}} \\[0.2em]
\texttt{**CRITICAL**: Always return the JSON object with the 'plagiarism\_segments' key. Do NOT return a bare empty array [] -- always wrap it in the object structure shown above.} \\[0.3em]
\texttt{\# Now, please process the following inputs:} \\[0.2em]
\texttt{<Paper\_A>} \\
\texttt{...full text...} \\
\texttt{</Paper\_A>} \\[0.2em]
\texttt{<Paper\_B>} \\
\texttt{...full text...} \\
\texttt{</Paper\_B>} \\
\bottomrule
\end{longtable}
\end{center}

\begin{center}
\small
\begin{longtable}{p{0.92\linewidth}}
\caption{Prompt template for contribution-level comparison in Phase~III.}
\label{tab:app_prompt_claim_comparison}\\
\toprule
\textbf{System Prompt:} \\
\texttt{You are a meticulous comparative reviewer for research papers. Your task is to determine whether a CANDIDATE paper refutes the novelty claims of an ORIGINAL paper.} \\[0.3em]
\texttt{**CRITICAL: CITATION REFERENCE USAGE**} \\
\texttt{When a citation reference (e.g., AgentGym-RL[1], Pilotrl[2]) is provided in the candidate paper title, you MUST use it throughout your analysis instead of phrases like 'The candidate paper'.} \\[0.3em]
\texttt{**TASK: Contribution Refutation Assessment**} \\
\texttt{For each contribution from the ORIGINAL paper, determine whether the CANDIDATE paper can refute the author's novelty claim (i.e., prove that the author was NOT the first to propose this contribution).} \\[0.3em]
\texttt{**REFUTATION STATUS DETERMINATION**:} \\
\texttt{1. **can\_refute**: The candidate demonstrates that similar prior work exists.} \\
\texttt{~~~- Provide detailed `refutation\_evidence` with summary (3-5 sentences) and evidence\_pairs.} \\
\texttt{~~~- Evidence pairs should show specific quotes from both papers that support this refutation.} \\
\texttt{~~~- If you observe that large portions of text are nearly identical, you should set this status.} \\
\texttt{2. **cannot\_refute**: The candidate does NOT challenge the novelty.} \\
\texttt{~~~- Provide ONLY a `brief\_note` (1-2 sentences) explaining technical differences.} \\
\texttt{~~~- Do NOT repeat the original paper's content. Be concise (e.g., 'This candidate focuses on web navigation tasks, not general RL frameworks.').} \\
\texttt{3. **unclear**: Comparison is difficult due to lack of detail.} \\[0.3em]
\texttt{**CRITICAL CONSTRAINTS**:} \\
\texttt{- Do NOT create artificial refutations by focusing on minor differences.} \\
\texttt{- Evidence quotes MUST be verbatim excerpts from the **Full Text Context** sections ($\leq$ 90 words per quote).} \\
\texttt{- For each quote, provide a `paragraph\_label` (e.g., 'Abstract', 'Introduction, Para 2', 'Methodology').} \\
\texttt{- Every quote MUST be found word-for-word in the provided context.} \\[0.3em]
\texttt{Output EXACTLY one JSON object:} \\[0.5em]

\textbf{User Prompt:} \\
\texttt{**Candidate Paper Title**: \{title\} (\{citation\})} \\
\texttt{**Number of Contributions to Compare**: \{N\}} \\
\texttt{**[Contributions to Compare]**} \\
\texttt{\{contributions\_text\}} \\
\texttt{**[Full Text Context: ORIGINAL]** (extract evidence from here)} \\
\texttt{```} \\
\texttt{\{orig\_text\}} \\
\texttt{```} \\
\texttt{**[Full Text Context: CANDIDATE]** (extract evidence from here)} \\
\texttt{```} \\
\texttt{\{cand\_text\}} \\
\texttt{```} \\
\texttt{**CRITICAL RULE**: You MUST extract quotes EXACTLY as they appear above. Copy character-by-character, including punctuation and spacing. If you cannot find a quote word-for-word in the context, do NOT use it. Evidence MUST be extracted from 'Full Text Context', NOT from Contribution Descriptions.} \\[0.3em]

\textbf{Expected Output Format (JSON):} \\
\texttt{\{"contribution\_analyses": [\{"aspect": "contribution", "contribution\_name": "...",} \\
\texttt{~"refutation\_status": "can\_refute" | "cannot\_refute" | "unclear",} \\
\texttt{~"refutation\_evidence": \{"summary": "3-5 sentences explaining HOW the candidate demonstrates prior work exists.",} \\
\texttt{~~"evidence\_pairs": [\{"original\_quote": "...", "original\_paragraph\_label": "...",} \\
\texttt{~~"candidate\_quote": "...", "candidate\_paragraph\_label": "...",} \\
\texttt{~~"rationale": "Explain how this pair supports refutation."\}, ...]\},} \\
\texttt{~"brief\_note": "1-2 sentences explaining why novelty is not challenged."\}, ...]\}} \\
\bottomrule
\end{longtable}
\end{center}

\begin{center}
\small
\begin{longtable}{p{0.95\linewidth}}
\caption{Prompt template for overall novelty assessment in Phase~III.}
\label{tab:app_prompt_overall_novelty_assessment}\\
\toprule
\textbf{System Prompt:} \\
\texttt{You are helping a human reviewer write the 'Originality / Novelty' assessment in a review form.} \\
\texttt{You are not deciding accept/reject; you are explaining how you read the paper's novelty in context.} \\[0.3em]
\texttt{You will receive FIVE kinds of signal:} \\
\texttt{1. The paper's abstract and a best-effort introduction (may be incomplete or imperfect).} \\
\texttt{2. A complete taxonomy tree showing the hierarchical structure of the research field (50 papers across \textasciitilde36 topics).} \\
\texttt{3. The original paper's position in this taxonomy (which leaf it belongs to, what sibling papers are in that leaf).} \\
\texttt{4. Literature search scope: how many candidate papers were examined (e.g., 30 papers from top-K semantic search).} \\
\texttt{5. Per-contribution analysis results with detailed statistics (e.g., Contribution A: examined 10 papers, 1 can refute).} \\[0.3em]
\texttt{CRITICAL CONTEXT:} \\
\texttt{- The taxonomy tree provides COMPLETE field structure: you can see how crowded or sparse each research direction is.} \\
\texttt{- The 'sibling papers' in the same taxonomy leaf as the original paper are the MOST relevant prior work.} \\
\texttt{- The statistics tell you the SCALE of the literature search (e.g., 30 candidates examined, not 300).} \\
\texttt{- 'is\_clearly\_refuted' ONLY means that among the LIMITED candidates examined, at least one appears to provide} \\
\texttt{~~~overlapping prior work. It does NOT mean the literature search was exhaustive.} \\
\texttt{- Use the taxonomy tree to assess whether the paper sits in a crowded vs. sparse research area.} \\[0.3em]
\texttt{Your job is to write 3--4 SHORT paragraphs that describe how novel the work feels given these signals.} \\
\texttt{Write as a careful, neutral reviewer. Each paragraph should focus on ONE specific aspect:} \\[0.3em]
\texttt{PARAGRAPH 1 (50--70 words): Core Contribution \& Taxonomy Position} \\
\texttt{- Summarize what the paper contributes.} \\
\texttt{- Use the taxonomy tree to explain where it sits: which leaf? How many sibling papers in that leaf?} \\
\texttt{- Is this a crowded or sparse research direction?} \\[0.3em]
\texttt{PARAGRAPH 2 (50--70 words): Field Context \& Neighboring Work} \\
\texttt{- Based on the taxonomy tree structure, identify nearby leaves and branches.} \\
\texttt{- Which related directions exist? How does this work connect to or diverge from them?} \\
\texttt{- Use the scope\_note and exclude\_note from taxonomy nodes to clarify boundaries.} \\[0.3em]
\texttt{PARAGRAPH 3 (50--70 words): Prior Work Overlap \& Literature Search Scope} \\
\texttt{- Discuss the contribution-level statistics: e.g., 'Contribution A examined 10 candidates, 1 can refute'.} \\
\texttt{- Be explicit about the search scale: 'Among 30 candidates examined...' not 'prior work shows...'.} \\
\texttt{- Which contributions appear more novel? Which have more substantial prior work?} \\[0.3em]
\texttt{PARAGRAPH 4 (40--60 words, OPTIONAL): Overall Assessment \& Limitations} \\
\texttt{- Brief synthesis of your impression given the LIMITED search scope.} \\
\texttt{- Acknowledge what the analysis covers and what it doesn't (e.g., 'based on top-30 semantic matches').} \\
\texttt{- Only include if you have substantive points; omit if redundant.} \\[0.3em]
\texttt{FORM REQUIREMENTS:} \\
\texttt{- Each paragraph must be a separate string in the output array (separated by \textbackslash n\textbackslash n when rendered).} \\
\texttt{- Do NOT use headings, bullet points, or numbered lists within paragraphs.} \\
\texttt{- Do not output numeric scores, probabilities, grades, or hard counts of papers/contributions.} \\
\texttt{- Avoid blunt verdicts ('clearly not novel', 'definitively incremental'); keep the tone descriptive and analytic.} \\
\texttt{- Write in English only.} \\[0.5em]
\textbf{User Prompt (JSON):} \\
\texttt{\{"paper\_context": \{"abstract": "...", "introduction": "..."\},} \\
\texttt{~"taxonomy\_tree": <full\_taxonomy\_json>,} \\
\texttt{~"original\_paper\_position": \{"leaf\_name": "...", "sibling\_count": N, "taxonomy\_path": [...]\},} \\
\texttt{~"literature\_search\_scope": \{"total\_candidates": N, "search\_method": "semantic\_top\_k"\},} \\
\texttt{~"contributions": [\{"name": "...", "candidates\_examined": N, "can\_refute\_count": M\}, ...],} \\
\texttt{~"taxonomy\_narrative": "..."\}} \\[0.3em]
\textbf{Expected Output Format (JSON):} \\
\texttt{\{"paragraphs": ["Paragraph 1 text...", "Paragraph 2 text...", "Paragraph 3 text...", "Paragraph 4 text (optional)"]\}} \\
\bottomrule
\end{longtable}
\end{center}

\begin{center}
\small
\begin{longtable}{p{0.92\linewidth}}
\caption{Prompt template for core-task sibling distinction in Phase~III.}
\label{tab:app_prompt_core_task_distinction}\\
\toprule
\textbf{System Prompt:} \\
\texttt{**CRITICAL: You MUST respond with valid JSON only. No markdown, no code fences, no explanations outside JSON.**} \\[0.3em]
\texttt{You are an expert in assessing research novelty within a specific research domain.} \\[0.3em]
\texttt{Your task is to compare the ORIGINAL paper against a CANDIDATE paper within the context of the core\_task domain, to assess the ORIGINAL paper's novelty.} \\[0.3em]
\texttt{Context Information:} \\
\texttt{- Core Task Domain: \{core\_task\_text\}} \\
\texttt{- Taxonomy Structure: The domain has been organized into a hierarchical taxonomy.} \\
\texttt{~~\{taxonomy\_context\_text\}} \\[0.3em]
\texttt{**RELATIONSHIP**: These papers are in the SAME taxonomy category (sibling papers).} \\
\texttt{They address very similar aspects of the core\_task.} \\[0.3em]
\texttt{**CRITICAL: DUPLICATE DETECTION**} \\
\texttt{First, determine if these papers are likely the same paper or different versions/variants:} \\
\texttt{- Check if the titles are extremely similar or identical} \\
\texttt{- Check if the abstracts describe essentially the same system/method/contribution} \\
\texttt{- Check if the core technical content and approach are nearly identical} \\[0.3em]
\texttt{If you determine they are likely duplicates/variants, output:} \\
\texttt{\{"is\_duplicate\_variant": true, "brief\_comparison": "This paper is highly similar to the original paper; it may be a variant or near-duplicate. Please manually verify."\}} \\[0.3em]
\texttt{If they are clearly different papers (despite being in the same category), provide a CONCISE comparison (2-3 sentences):} \\
\texttt{\{"is\_duplicate\_variant": false, "brief\_comparison": "2-3 sentences covering: (1) how they belong to the same parent category, (2) overlapping areas with the original paper regarding the core\_task, (3) key differences from the original paper."\}} \\[0.3em]
\texttt{**Requirements for brief\_comparison (when is\_duplicate\_variant=false)**:} \\
\texttt{- Sentence 1: Explain how both papers belong to the same taxonomy category (shared focus/approach)} \\
\texttt{- Sentence 2-3: Describe the overlapping areas and the key differences from the original paper} \\
\texttt{- Be CONCISE: 2-3 sentences ONLY} \\
\texttt{- Focus on: shared parent category, overlapping areas, and key distinctions} \\
\texttt{- Do NOT include quotes or detailed evidence} \\
\texttt{- Do NOT repeat extensive details from the original paper} \\[0.3em]
\texttt{**Output format (STRICT JSON only, no markdown, no code fences, no extra text)**:} \\
\texttt{\{"is\_duplicate\_variant": true/false, "brief\_comparison": "2-3 sentences as described above"\}} \\[0.3em]
\texttt{IMPORTANT: Your ENTIRE response must be valid JSON starting with \{ and ending with \}.} \\
\texttt{Do NOT include any text before or after the JSON object.} \\[0.5em]
\textbf{User Prompt (JSON):} \\
\texttt{\{"core\_task\_domain": "...", "original\_paper": \{"title": "...", "content": "...",} \\
\texttt{~"content\_type": "fulltext|abstract"\}, "candidate\_paper": \{"title": "...", "content": "...",} \\
\texttt{~"content\_type": "fulltext|abstract"\}, "analysis\_instruction": "These papers are classified in the SAME leaf category in the taxonomy (sibling papers). First check if they are likely duplicates/variants by comparing titles, authors, and content. If not duplicates, provide a concise 2-3 sentence comparison covering: (1) their shared taxonomy position, (2) overlapping areas with original paper, (3) key differences."\}} \\[0.3em]
\textbf{Expected Output Format (JSON):} \\
\texttt{\{"is\_duplicate\_variant": true|false, "brief\_comparison": "..."\}} \\
\bottomrule
\end{longtable}
\end{center}

\begin{center}
\small
\begin{longtable}{p{0.92\linewidth}}
\caption{Prompt template for subtopic comparison in Phase~III.}
\label{tab:app_prompt_subtopic_comparison}\\
\toprule
\textbf{System Prompt:} \\
\texttt{You compare an original taxonomy leaf against sibling subtopics to summarize similarities and differences.} \\
\texttt{Focus on category boundaries and representative papers; do NOT invent details.} \\[0.5em]
\midrule
\textbf{User Prompt (JSON):} \\
\texttt{\{"core\_task": "<core\_task\_text>",} \\
\texttt{~"original\_leaf": \{"name": "...", "scope\_note": "...", "exclude\_note": "...", "paper\_ids": [...]\},} \\
\texttt{~"sibling\_subtopics": [\{"name": "...", "scope\_note": "...", "exclude\_note": "...",} \\
\texttt{~~~"leaf\_count": N, "paper\_count": M, "papers": [\{"title": "...", "abstract": "...", "id": "..."\}, ...]\}, ...],} \\
\texttt{~"instructions": "Return concise JSON with keys: overall (2--4 sentences), similarities (list),} \\
\texttt{~~~differences (list), suggested\_search\_directions (optional list).} \\
\texttt{~~~Base reasoning on scope/exclude notes and the provided subtopic papers (titles/abstracts).} \\
\texttt{~~~Do not invent facts."\}} \\[0.3em]
\textbf{Expected Output Format (JSON):} \\
\texttt{\{"overall": "2--4 sentence summary of relationship between original leaf and sibling subtopics",} \\
\texttt{~"similarities": ["similarity 1", "similarity 2", ...],} \\
\texttt{~"differences": ["difference 1", "difference 2", ...],} \\
\texttt{~"suggested\_search\_directions": ["direction 1", ...] (optional)\}} \\
\bottomrule
\end{longtable}
\end{center}

\FloatBarrier

\subsection{Phase~III Complete Report Schema}
\label{app:phase3_report_schema}

Phase~III produces a comprehensive JSON report (\texttt{phase3\_complete\_report.json}) that serves as the sole input for Phase~IV rendering. Table~\ref{tab:app_report_modules} summarizes the seven modules of this report.

\begin{table}[ht]
\centering
\caption{Phase~III complete report module definitions}
\label{tab:app_report_modules}
\small
\setlength{\tabcolsep}{0.5em}
\begin{tabular}{lp{7cm}}
\toprule
\textbf{Module} & \textbf{Contents} \\
\midrule
\texttt{original\_paper} & Title, authors, abstract, venue, PDF URL (from Phase~I) \\
\texttt{core\_task\_survey} & Taxonomy tree, narrative synthesis (2 paragraphs), papers index, display metadata \\
\texttt{contribution\_analysis} & Overall novelty assessment (3--4 paragraphs), per-contribution comparisons with verified evidence, statistics \\
\texttt{core\_task\_comparisons} & Sibling paper comparisons, structural position info \\
\texttt{references} & Citation index with aliases and URLs (from Phase~II) \\
\texttt{textual\_similarity} & Unified textual similarity index, segments by candidate paper \\
\texttt{metadata} & Generation timestamp, pipeline version, component flags, artifact filenames \\
\bottomrule
\end{tabular}
\end{table}

\subsection{Refutation Entry Schema}
\label{app:refutation_schema}

Each per-contribution comparison entry contains the following fields:

\begin{itemize}
    \item \texttt{refutation\_status}: One of \texttt{can\_refute}, \texttt{cannot\_refute}, or \texttt{unclear}
    \item \texttt{refutation\_evidence} (if \texttt{can\_refute}): Object containing:
    \begin{itemize}
        \item \texttt{summary}: 3--5 sentence explanation of how prior work challenges the claim
        \item \texttt{evidence\_pairs}: List of evidence pairs, each containing:
        \begin{itemize}
            \item \texttt{original\_quote}: Verbatim excerpt from target paper ($\leq$90 words)
            \item \texttt{original\_paragraph\_label}: Location label (e.g., ``Methods'')
            \item \texttt{original\_location}: Verification result object:
            \begin{itemize}
                \item \texttt{found}: Boolean indicating if quote was verified in source
                \item \texttt{match\_score}: Confidence score from verification algorithm (0.0--1.0)
            \end{itemize}
            \item \texttt{candidate\_quote}: Verbatim excerpt from candidate paper ($\leq$90 words)
            \item \texttt{candidate\_paragraph\_label}: Location label
            \item \texttt{candidate\_location}: Verification result object (same structure as \texttt{original\_location})
            \item \texttt{rationale}: Brief explanation of how this pair supports refutation
        \end{itemize}
    \end{itemize}
    \item \texttt{brief\_note} (if not \texttt{can\_refute}): 1--2 sentence explanation
\end{itemize}
The \texttt{unclear} judgment is used when the relationship cannot be determined due to insufficient information or ambiguous overlap. These cases receive only a brief explanatory note and are treated identically to \texttt{cannot\_refute} in statistics (counted as \texttt{non\_refutable\_or\_unclear}) and binary novelty classification.

\FloatBarrier

\section{Phase~IV: Report Rendering}
\label{app:phase4}

Phase~IV performs template-based rendering of the novelty assessment report. This phase involves \textbf{no LLM calls}; all content is assembled from the Phase~III complete report.

\subsection{Rendering Pipeline}
\label{app:phase4_pipeline}

The rendering pipeline consists of three stages:

\paragraph{Input.}
Phase~IV reads the \texttt{phase3\_complete\_report.json} file produced by Phase~III (schema defined in Appendix~\ref{app:phase3_report_schema}).

\paragraph{Template Assembly.}
The report generator constructs a Markdown document by iterating through the seven modules and applying section-specific templates. Key rendering decisions include:
\begin{itemize}
    \item \textbf{Quote truncation}: Evidence quotes exceeding display limits are truncated with ellipsis markers
    \item \textbf{Citation formatting}: References are rendered using the unified citation index from Phase~II
    \item \textbf{Taxonomy visualization}: The hierarchical taxonomy is rendered as an indented text tree
\end{itemize}

\paragraph{Output Conversion.}
The Markdown report is optionally converted to PDF using Pandoc with custom styling. Output filenames are deterministically generated from Phase~III metadata to ensure reproducibility.

\subsection{Report Structure}
\label{app:report_structure}

The rendered Markdown report contains the following sections:
\begin{enumerate}
    \item \textbf{Header}: Paper title, PDF URL, generation metadata
    \item \textbf{Core Task Survey}: Taxonomy tree and 2-paragraph narrative synthesis
    \item \textbf{Core Task Comparisons}: Sibling paper distinctions (if applicable)
    \item \textbf{Contribution Analysis}: Per-contribution comparisons with overall novelty assessment
    \item \textbf{Textual Similarity} (Appendix): Detected overlap segments (if any)
    \item \textbf{References}: Unified citation list
\end{enumerate}

\end{document}